\documentclass[12pt]{article}
\usepackage{amsmath}
\usepackage{graphicx,psfrag,epsf}
\usepackage{enumerate}
\usepackage{natbib}
\usepackage{url} 

\newcommand{\blind}{1}

\renewcommand{\hat}{\widehat}
\def\wh{\widehat}
\def\wt{\widetilde}
\def\askip{\vspace{0.1in}}

\newcommand{\IID}{{\rm IID}}

\newcommand{\cov}{{\rm Cov}}

\newcommand{\var}{{\rm Var}}

\def\la{\lambda}

\newcommand{\ve}{{\varepsilon}}

\newcommand{\bSigma}{\boldsymbol{\Sigma}}

\newcommand{\bve}{\mbox{\boldmath$\varepsilon$}}

\newcommand{\calI}{{\mathcal I}}

\newcommand{\calS}{{\mathcal S}}

\newcommand{\calB}{{\mathcal B}}
\newcommand{\calC}{{\mathcal C}}
\newcommand{\calE}{{\mathcal E}}

\def\6bullets{\bullet\bullet\bullet\bullet\bullet\bullet}

\usepackage{geometry}
\usepackage{color}
\usepackage{rotating}
\usepackage{threeparttable}
\usepackage{multirow}
\usepackage{subcaption}

\begin{document}

\def\spacingset#1{\renewcommand{\baselinestretch}%
	{#1}\small\normalsize} \spacingset{1}

\if1\blind
{
  \title{\large\bf 
Probabilistic Forecasting for Daily Electricity Loads\\ and
Quantiles for Curve-to-Curve
Regression\thanks{Ying Chen acknowledges the  support from the Singapore
Ministry of
Education Academic Research Fund Tier 1 at National University of Singapore.
Qiwei Yao acknowledges the support from the Fondation Math{\'e}matique
Jacques Hadamard. }}
  \author{\normalsize Xiuqin Xu \\
  	\normalsize NUS Graduate School for Integrative Sciences
  	\& Engineering and  Institute of \\
  	 \normalsize Data  Science,  National University of Singapore, Singapore, xiuqin.xu@u.nus.edu
    \and \normalsize  Ying Chen  \\ \normalsize
    Department of Mathematics \& Risk Management Institute\\  \normalsize
    National University of Singapore, Singapore,
    matcheny@nus.edu.sg 
    \and \normalsize Yannig Goude \\ \normalsize
    \'Electicit\'e de France R\&D  \& Laboratoire de Math\'ematique
d'Orsay, Paris, France\\ 
  \normalsize  yannig.goude@edf.fr
    \and  \normalsize Qiwei Yao \\
  \normalsize   Department of Statistics, London School of Economics and Political Science,\\
  \normalsize London, United Kingdom, q.yao@lse.ac.uk}
  \maketitle
} \fi

\if0\blind
{
  \bigskip
  \bigskip
  \bigskip
  \begin{center}
    {\large\bf
Probabilistic Forecasting for Daily Electricity Loads\\ and
Quantiles for Curve-to-Curve
Regression}
\end{center}
  \medskip
} \fi

\begin{abstract}
Probabilistic forecasting of electricity load curves is of fundamental importance for effective scheduling and decision making in the increasingly volatile and competitive energy markets.  We propose a novel approach to construct probabilistic predictors for curves (PPC), which leads to a natural and new definition of quantiles in the context of curve-to-curve linear regression. There are three types of PPC: a predictive set, a predictive band and a predictive quantile,  all of which are defined at a pre-specified nominal probability level. In the simulation study, the PPC achieve promising coverage probabilities under a variety of data generating mechanisms. When applying to one day ahead forecasting for the French daily electricity load curves, PPC outperform several state-of-the-art predictive methods in terms of forecasting accuracy, coverage rate and average length of the predictive bands. 
The predictive quantile curves provide insightful information which is highly relevant to hedging risks in electricity supply management.
\end{abstract}

\noindent%


\spacingset{1.45} 

\section{Introduction}


Electricity load forecasting is an essential element for effective
scheduling and decision making in energy markets. The conventional
methods of load forecasting are fundamentally deterministic and focus on
the mean level of future consumption. However, the future is uncertain and
costs are more driven by extreme events when, for example,  the electricity storage
capacities are exhausted. The modern upgrades of power grids with the
integration of renewable energy and development of efficient electricity
management systems further introduce uncertainty and fluctuations in the
volatile markets. This advocates for probabilistic forecasting methods,
aiming at not only accurate point prediction of load, but also insightful
predictive intervals and predictive quantiles at pre-determined nominal
probability levels.  

One-day-ahead high temporal resolution (hourly or half-hourly) load forecasting
plays a key role in power system planning and operation, where many
operating decisions rely on the load forecasts, such as dispatch
scheduling of production transformation, reliability analysis and demand
management, see \cite{rolnick2019tackling}. On one hand, the development
and integration of renewable production such as photovoltaic panel or
wind turbines increases the proportion of electricity production units
dependent on meteorological conditions, making the supply of electricity
more volatile and unpredictable, see \cite{GIELEN201938}.  On the other
hand, the electricity storage capacities are still expensive and limited,
though the smart grid infrastructures  \citep{Wang18} and smart charging programs \citep{garcia2014plug} allow
for more information to transit between end users and producers,
increasing the efficiency of demand response \citep{Wang15}.

While a point prediction is most frequently used in forecasting  future
electricity loads, the associated risk and the uncertainty are not
clear. \cite{HONG2016914} survey the available models/methods
for probabilistic forecasting for electricity loads, including evaluation methods
and common misunderstanding. See also \cite{hong2016probabilistic, HONG20191389}. A predictive interval with a pre-specified coverage probability is
more informative, which
is arguably the most frequently used probabilistic forecaster. Nevertheless, most available methods forecast individual
loads separately, as simultaneous probabilistic forecasting for multiple values 
(such as the loads on each 30-minute interval over a day) imposes extra 
complication and challenges. Simply merging these individual predictive intervals for multiple loads
loses the probability interpretation immediately. There is intricate dependence among different predictive intervals, making it  formidable to adjust each individual
coverage probability. The
Bonferroni correction is often too crude to be useful for approximation. As mentioned in \cite{Polonik2000}, 
direct construction of a joint predictive region, with a pre-specified 
coverage probability,
is only possible in some simple cases.
Even then, one faces the difficulties
in choosing the geometric shape of the region.


The high temporal resolution of daily loads makes it attractive to review the loads over a day as a curve. In the functional space, it embeds non-stationary daily patterns into a stationary framework in a Hilbert
space, and has provided competitive and reliable
pointwise forecasting;  see \cite{Cho2013,Cho2015,chen2017adaptive,chen2020day}. There is rich literature on interval forecasting of electricity loads. \cite{Taylor2003} constructed interval forecasting based on weather ensemble prediction consisting of 51 weather scenarios. 
\cite{Petiau2009} proposed interval forecasts based on empirical quantiles of the relative forecasting errors in the past.  
\cite{Kou2014} proposed a heteroscedastic Gaussian model to predict the distributions of one day ahead electricity loads.
Recently, there are some approaches extending quantile regression based on pin-ball loss optimization of \cite{koenker1978regression} to produce interval forecasts, such as quantile additive models \citep{gaillard2016additive,dordonnat2016gefcom2014,fasiolo2020fast}, tree-based ensemble with gradient boosting models \citep{roach2019reconciled}, and ensemble of experts neural network, quantile random forest and tree-based ensemble \citep{smyl2019machine}.
Apart from constructing predictive intervals,  \cite{Cabrera2017} derived
daily quantile curves based on pointwise quantile estimation and
forecasted the future load curves based on functional principal component
analysis. Unfortunately, the quantile curves do not inherent the
probability interpretation of pointwise quantiles. Therefore, the
coverage probabilities of the resulting forecasting bands are unknown, as the above mentioned other works.
\cite{antoniadis2016prediction} proposed a nonparametric function-valued
model which combines kernel regression and wavelet transformation to
produce simultaneous loads predictions at multiple time horizons, where
 construction of predictive interval for the whole daily curves is
considered difficult. 

In the context of curve regression, our objective is to develop some probabilistic predictors for
curves (PPC) with a pre-specified coverage probability. In particular, we advocate three types
of PPC: a predictive set which consists of a bundle of curves, a predictive
band which is a continuous region, and a predictive quantile which labels the outside curve.
The new approach is to transfer a curve-to-curve
linear regression into several scalar linear regressions, where we construct a joint predictive region for the error terms across all scalar
linear regression models and induce a predictive set for the original targeted curve. 
 The calibration of the nominal coverage probability is achieved via either
a residual-based $\chi^2$-approximation or the empirical distributions
of the residuals. There is a challenge that real-world data only have finite sample size, and often small sample size, for which the asymptotic theory fails to provide insightful advice on the construction of probabilistic forecasting. Thus, a resampling method is proposed for the latter. 
We define the envelope of the predictive set as the predictive band, and the most
``outside'' curves in the predictive set as the predictive quantiles using the concept of the extremal depth for curve data
\citep{narisetty2016extremal}.


To our best knowledge, this is the first attempt to construct probabilistic
forecasting predictors at a given nominal probability for a curve.  The proposed 
quantile curves also admit natural and explicit probability interpretation.  Simulation study indicates that PPC achieve accurate coverage rates under various scenarios. 
When applying to the French electricity loads data with a temporal resolution of 30 minutes,
PPC directly provide the probabilistic forecasts for the daily loads as a whole.
The predictive quantiles at different probability levels deliver insightful
information on prospective future scenarios, which is valuable for hedging
risks in electricity management. 
It also outperforms several state-of-the-art  methods in terms
of not only more accurate pointwise forecasts, but also more accurate predictive intervals
with more accurate coverage probability and shorter length of the intervals. The code and data are available in GitHub\if1\blind{: \url{https://github.com/Sherry-Xu/Probabilistic-Forecasting-for-Curves}\fi \if0\blind { (the link is removed for the blinded version)}\fi.

The rest of the article is structured as follows. Section
\ref{sec:method} introduces the curve linear regression framework 
and proposes the method to
construct PPC, including predictive sets, predictive bands and the predictive quantiles
for curves. Simulation studies investigating the finite sample performance of the
proposed PPC methodology are conducted in Section
\ref{sec:simulation}. In Section \ref{sec:realdata}, PPC are applied to
predict day-ahead French electricity load curves in 2019 based on the
historical values from 2012 to 2018. Section \ref{sec:conclusion} concludes. 

\section{Methodology} \label{sec:method}

\subsection{Curve regression and dimension reduction}
Let $Y_t(u), u \in \calI_1$, be the electricity load curve on the $t$-th day. Associated
with each $Y_t(\cdot)$, there is a regressor curve $X_t(v), v \in \calI_2$, which 
may be $Y_{t-1}(\cdot)$, or multiple lagged curves such as $Y_{t-1}(\cdot)$ and $Y_{t-2}(\cdot)$ together, or even contain multiple
exogenous variables such as (predicted) temperature curve for the day.
We assume that the first two moments of $\{ Y_t(\cdot), X_t(\cdot) \}$ are time-invariant.

Consider the curve linear regression, see \cite{Cho2013,Cho2015}:
\begin{equation} \label{b1}
Y_t(u) = \mu_{y}(u) + \int_{\calI_2} \{ X_t(v) - \mu_{x}(v) \} \beta (u, v) dv + \ve_t(u),
\quad u \in \calI_1,
\end{equation}
where $\mu_{y}(u) =E\{Y_t(u) \}, \; \mu_{x}(v) =E\{X_t(v) \}$, $\ve_t(\cdot)$ is zero-mean
independent noise curve. 
Perform the singular-value-decomposition (SVD):
\[
\Sigma_{yx}(u,v) \equiv \cov( Y_t (u), X_t(v) ) = \sum_{j=1}^\infty
\sqrt{\la_{j}} \varphi_{j}(u) \psi_{j}(v),  
\]
where $\la_{1} \ge \la_2 \ge \cdots \ge 0 $ are singular values. It holds  that
\begin{equation} \label{b0}
Y_t(u) - \mu_y(u) = \sum_{j=1}^{\infty} \xi_{tj} \varphi_{j}(u), \quad
X_t(v) - \mu_x(v)  = \sum_{j=1}^{\infty} \eta_{tj} \psi_{j}(v),
\end{equation}
where 
\[
\xi_{tj} = \int_{\calI_1} \{ Y_t(u) - \mu_y(u) \} \varphi_{j}(u) du, \quad
\eta_{tj} = \int_{\calI_2} \{ X_t(u) - \mu_x(u) \} \psi_{j}(u) du.
\]
It follows from Theorem 1 of \cite{Cho2013} that the curve
regression (\ref{b1}) is 
equivalent to
\begin{equation} \label{b3}
\xi_{tj} = \sum_{\ell=1}^\infty b_{j \ell} \eta_{t\ell} + \varepsilon_{tj}, \quad
\ve_{tj} \sim (0, \sigma_j^2), \quad
j=1, 2, \cdots, 
\end{equation}
where 
\[
b_{j\ell} = \int_{\calI_1 \times \calI_2} \varphi_{j}(u)\psi_{\ell}(v) \beta(u,v) dudv,
\quad 
\varepsilon_{tj} = \int_{\calI_1} \varepsilon_t(u) \varphi_{j}(u) du.
\]

To simplify the exploration, we assume from now on
\begin{equation} \label{b00}
Y_t(u) - \mu_y(u) = \sum_{j=1}^d \xi_{tj} \varphi_j(u),
\end{equation}
where $d\ge 1$ is an unknown but finite integer.
We will specify how to estimate $d$ below. 
Furthermore, we assume that (\ref{b3}) admits the finite expression
\begin{equation} \label{b33}
\xi_{tj} = \sum_{\ell\in \pi_j} b_{j \ell} \eta_{t\ell} + \varepsilon_{tj}, \quad
\ve_{tj} \sim (0, \sigma_j^2), \quad
j=1,  \cdots, d,
\end{equation}
where $\pi_j$ is a set containing the indices of the finite number of the
 regressors $\eta_{t\ell}$ for $\xi_{tj}$. By the virtue of  SVD, it holds that $j \in \pi_j$.
Theoretically, one could pursue more complete approach by assuming $d=\infty$
and then truncate the sums in (\ref{b00}) and (\ref{b33}) by some
asymptotic approximations. This, however, has little
bearing in terms of applications given that many real curve data from e.g.,
finance and energy sectors do exhibit finite dimensional behavior.

With available data $\{ ( Y_t(\cdot), X_t(\cdot)), \; 1\le t \le N\}$,
put
\begin{equation} \label{a2}
\wh \mu_y(u) = {1 \over N} \sum_{t=1}^N Y_t(u) , \quad \qquad
\wh \mu_x(u) = {1 \over N} \sum_{t=1}^N X_t(u),
\end{equation}
\[
\wh \Sigma_{yx}(u,v) = {1 \over N} \sum_{t=1}^N \{ Y_t(u) - \wh \mu_y(u)
\} \{ X_t(v) - \wh \mu_x(v) \}.
\]
Performing SVD on $\wh \Sigma_{yx}(u,v)$, we obtain
\begin{equation} \label{a3}
\wh \Sigma_{yx}(u,v) =  \sum_{j=1}^{\infty }
\wh \la_{j}^{1\over 2}  \wh \varphi_{j}(u)  \wh \psi_{j}(v), 
\end{equation}
where $\wh \la_1 \ge \wh \la_2 \ge \cdots $ are the singular values of
$\wh \Sigma_{yx}(u,v)$.
Now 
replacing $\{ \xi_{tj}, \eta_{tj} \}$ in (\ref{b33}) by
\begin{equation} \label{a4}
\wh \xi_{tj} = \int_{\calI_1} \{ Y_t(u) - \wh \mu_y(u) \} \wh \varphi_{j}(u) du \quad
{\rm and} \quad
\wh \eta_{tj} = \int_{\calI_2} \{ X_t(u) - \wh \mu_x(u) \} \wh \psi_{j}(u) du.
\end{equation}

Given the large number of parameters, we select regressors 
for each fixed $j$ using stepwise regression controlled by AIC. This leads to an estimated index set $\wh\pi_j$. Other methods such as the regularized least squares estimation with  $\ell_1$ (Lasso) or $\ell_2$ (Ridge) penalty can also be used, which produce similar performance in our analysis and are omitted in the manuscript. The fitted model is then of the form
\begin{equation}
\label{equ:1}
\wh \xi_{tj} = \sum_{\ell\in \wh\pi_j } \hat{b}_{j \ell} \wh \eta_{t\ell} + \wh \ve_{tj} , \quad
j=1, \cdots, \wh d, 
\end{equation}
where $\wh d$ is an estimator for $d$ to be specified below,
and
\begin{equation} \label{b14}
\wh \ve_{tj} = \wh \xi_{tj} - \sum_{\ell \in \wh\pi_j} \wh b_{j\ell}
\wh \eta_{t\ell}, 
\qquad \quad
\wh \sigma_j^2 = {1 \over N- |\wh \pi_j|} \sum_{t=1}^N \big( \wh \ve_{tj} 
\big)^2.
\end{equation}
In the above expression, $|\wh \pi_j|$ denotes the cardinality of $\wh \pi_j$.

When estimating the predictive curve $E\{ Y_t(\cdot) | X_t(\cdot) \}$, \cite{Cho2013} 
chose $d$ by
\[
\wt d_1 = \arg \min_{1\le j \le d_0 } \wh \la_{j+1}\big/ \wh \la_{j},
\]
where $d_0 $ is a pre-specified positive integer. 
Given that $\la_1 \ge \cdots \ge \la_d >0 = 
\la_{d+1} = \cdots $, this selection attempts to ensure that $\xi_{t1}, \cdots, \xi_{td}$ catch
all the information on $Y_t(\cdot)$ from $X_t(\cdot)$.
Although $\wt d_1$ is appropriate for constructing a confidence set
for the expectation curve $E\{ Y_t(\cdot) | X_t(\cdot) \}$, it is
different from our goal of constructing a predictive set for
$Y_t(\cdot)$
given $X_t(\cdot)$, for which the noise term $\ve_t(\cdot)$ in (\ref{b1}) also matters.
Let $\wt d_2$ be the minimum value of $d$
such that the total variation
of  the RHS of (\ref{b00}) over $t=1, \cdots, N$ accounts for more than a
certain threshold, e.g.  99.9\% of total variation of $\{
Y_t(\cdot) \}$, 
we estimate $d$ by 
\begin{equation}
\hat{d} =  \max(\wt d_1, \wt d_2).
\label{a8}
\end{equation}

\subsection{Predictive sets for $Y(\cdot)$ given $X(\cdot)$}

Given a new value of $X(\cdot)$, our goal is to predict $Y(\cdot)$ defined
by the RHS of (\ref{b1}) with $\{X_t(\cdot), \ve_t(\cdot)\}$ replaced
by $\{X(\cdot), \ve(\cdot)\}$, where $\ve(\cdot)$ is unobservable.
Then it follows from (\ref{b0}) and (\ref{b00}) that
\begin{equation} \label{b02}
Y(u) - \mu_y(u) = \sum_{j=1}^{d} \xi_{j} \varphi_{j}(u), \quad
X(v) - \mu_x(v)  = \sum_{j=1}^{\infty} \eta_{j} \psi_{j}(v).
\end{equation}

Denote $\bve_t(d) = (\ve_{t1}, \cdots, \ve_{td})'$ and $\bSigma = \var\{\bve_t(d)\}$.
Write $\bve(d) = (\ve_1, \cdots, \ve_d)'$.
For any $\alpha \in (0, 1)$, define
\[
\calE_{1 -\alpha} =\big\{ \bve(d) \, \big| \,
\bve(d)' \bSigma^{-1} \bve(d)
\le C_{ \alpha,d}   \big\},
\]
where $0  < C_{ \alpha,d} < \infty$ is a constant determined by
\begin{equation} \label{a5}
P\{ \bve_t(d) \in \calE_{1-\alpha} \} = 1- \alpha.
\end{equation}
Put
\[
\calC_{1- \alpha} (X)  = \Big\{ \mu_y(\cdot) + \sum_{j=1}^d \xi_{j} \varphi_j(\cdot) \,
\Big| 
\Big(\xi_1 -  \sum_{\ell\in \pi_1} b_{1 \ell} \eta_{\ell} ,\, \cdots,\,
\xi_d -   \sum_{\ell\in\pi_d} b_{d \ell} \eta_{\ell} \Big)' \in \calE_{1-\alpha} \Big\}.
\]
It follows from (\ref{b33}), (\ref{b02}) and (\ref{a5})   that
\[
P\{ Y (\cdot) \in \calC_{1-\alpha} (X) | X(\cdot) \} = 1-\alpha,
\]
i.e. $\calC_{1-\alpha} (X)$ is a true predictive set for $Y(\cdot)$ based on $X(\cdot)$
with the nominal coverage probability $1-\alpha$.
In practice, we replace $d$ by $\wh d$ in (\ref{a8}), $\mu_y(\cdot)$ by $\wh \mu_y(\cdot)$
in (\ref{a2}), $\varphi_j(\cdot)$ by $\wh \varphi_j(\cdot)$ in (\ref{a3}),
and
$(b_{j\ell}, \pi_j)$ by $(\wh b_{j\ell}, \wh \pi_j)$ in (\ref{equ:1}).
A 
data based predictive set, i.e. an estimator for $\calC_{1-\alpha} (X)$ can be
defined as 
\begin{equation} \label{b15}
\wh \calC_{1-\alpha}(X) =
\Big\{\, \wh \mu_y(\cdot) + \sum_{j=1}^{\wh d} \Big( \ve_{j} +
\sum_{\ell\in \wh \pi_j} \wh b_{j\ell} \wh
\eta_{\ell}\Big) \wh \varphi_j(\cdot) \,
\Big| \,
\bve(\wh d)' \wh{\bSigma}^{-1} \bve(\wh d) 
\le C_{ \alpha, \wh d}  
 \, \Big\},
\end{equation}
where $\{\wh \eta_{\ell}\}$ are obtained in (\ref{a4}) with
$X_t(\cdot)$ replaced by $X(\cdot)$, 
$\wh {\bSigma}$ is the sample covariance matrix of $\wh {\bve}_t
\equiv (\wh \ve_{t1}, \cdots, \wh \ve_{t \wh d})'$
for $t=1, \cdots, N$, and $\wh \ve_{t j}$ is given in (\ref{b14}). 
The $(i,j)$-th element of $\wh {\bSigma}$ is defined as
$ \sum_{t=1}^N\wh \ve_{ti}\wh \ve_{tj}\big/(N - \gamma_{ij})$, where
$\gamma_{ij}$ is the  cardinality of $\wh \pi_i \cup  \wh \pi_j$.
We propose two ways to determine the constant $C_{ \alpha, \wh d}$\,.
\begin{itemize}
\item[(i)] $\chi^2$-approximation:  Assuming that $\ve_{t1},  \cdots, \ve_{td}$ are jointly normal,
then \linebreak $\bve_t(d)' \bSigma^{-1} \bve_t(d) \sim \chi^2_d$\,.
Let
$C_{\alpha, \wh d}$
be the $(1-\alpha)$-th percentile of the $\chi^2$ distribution
with $\wh d$ degrees of freedom.
We use in (\ref{b15})
\begin{equation} \label{b17}
(\ve_1, \cdots, \ve_{\wh d})' = \wh {\bSigma}^{1/2} (z_1, \cdots, z_{\wh d})', 
\end{equation}
where $z_1, \cdots, z_{\wh d}$ are independent and follow $\mathcal{N} (0,1)$,
and $\sum_i
z_i^2\le C_{\alpha, \wh d}.$
In principle, $\wh \calC_{1-\alpha} (X)$ consists of
infinite number of curves.
\item[(ii)] Empirical distribution for residuals:
Let  $C_{\alpha, \wh d}$ be the  $(1-\alpha)$-th percentiles
of the empirical distribution of
\begin{equation} \label{d3}
(\wh \ve_{t1}, \cdots, \wh \ve_{t \wh d})'\,\wh {\bSigma}^{-1} (\wh \ve_{t1},
\cdots, \wh \ve_{t \wh d})
\qquad t=1, \cdots, N.
\end{equation}
Then $\wh \calC_{1-\alpha} (X)$ consists of finite number of curves, i.e.
$[N(1-\alpha)]$ curves generated via $\bve_t(\wh d) 
=(\wh \ve_{t1}, \cdots, \wh \ve_{t\wh d})'$ for which the inequality in (\ref{b15})
holds and $1 \le t \le N$.
\end{itemize}

\subsection{Predictive bands for $Y(\cdot)$ given $X(\cdot)$}
\label{subsec:bands}

In spite of the clear probability interpretation, the predictive set $\wh
\calC_{1-\alpha(X)}$ consists of a bundle of curves. In practice, it is
more convenient to use a band or a region which covers the target curve
$Y(\cdot)$ with probability $1 - \alpha$. A natural predictive
band is the envelope  of  $\wh \calC_{1-\alpha}(X)$:
\[
\wh\calB_{1-\alpha}(X) = \big(\wh Y_{\rm low}(\cdot), \;\wh Y_{\rm up}(\cdot) \big), 
\]
where 
\begin{equation} \label{b19}
\wh Y_{\rm low}(u) = \min \{ Z(u)\, \big|\,  Z(\cdot) \in \wh \calC_{1-\alpha}(X) \}, \quad
\wh Y_{\rm up}(u) = \max \{ Z(u)\, \big|\,  Z(\cdot) \in \wh \calC_{1-\alpha}(X) \}. 
\end{equation}

For $\wh \calC_{1-\alpha}(X)$ constructed based on the $\chi^2$ distribution,
the infinite number of curves in $\wh \calC_{1-\alpha}(X)$ fill in every
space in $\wh\calB_{1-\alpha}(X)$
due to the continuity of the normal distribution; see also
(\ref{b17}). It holds that
\begin{equation} \label{b20}
P\{ \, Y(\cdot) \in \wh\calB_{1-\alpha}(X) \, |\, X \,\} \;\ge \;
P\{ \, Y(\cdot) \in \wh\calC_{1-\alpha}(X) \, |\, X \,\} \; \approx \; 1 - \alpha.
\end{equation}
As an alternative, the empirical distribution based approach is robust when the stochastic noise deviates from the normal distribution. 
On the other hand, there is a potential problem in $\wh\calB_{1-\alpha}(X)$
when the confidence set $\wh\calC_{1-\alpha}(X)$ is constructed based on the empirical
distribution, as then $\wh\calC_{1-\alpha}(X)$ consists of merely
$[N(1-\alpha)]$ curves, and
the width $\wh Y_{\rm up}(\cdot) - \wh Y_{\rm low}(\cdot)$ increases
as $N$ increases. For small $N$, the band $\wh\calB_{1-\alpha}(X)$ could
be too narrow to cover $Y(\cdot)$ with probability $1-\alpha$. For
forecasting electricity load curves as well as other real world data, the
sample sizes are often small in order to retain the ``stationarity'' required in the curve regression model.

We thus propose a resampling method to adjust the coverage
probability of $\wh\calB_{1-\alpha}(X)$.
 For a given positive integer $K$, we draw $\ve_{1j}^\star, \cdots,
\ve_{Kj}^\star$ independently with replacement from
$\wh\ve_{1j}, \cdots, \wh\ve_{Nj}$, $j=1, \cdots, \wh d$. Put
$\bve_i^\star =(\ve_{i1}^\star, \cdots, \ve_{i\wh d}^\star)'$ and \small{
\begin{equation} \label{b18}
\calC^\star_{1-\alpha}(X) = 
\Big\{\, \wh \mu_y(\cdot) + \sum_{j=1}^{\wh d} \Big(\ve_{ij}^\star + 
\sum_{\ell \in \wh \pi_j} \wh b_{j\ell} \wh
\eta_{\ell}\Big) \wh \varphi_j(\cdot) \,
\Big| \,
(\bve_i^\star)' \wh{\bSigma}^{-1} \bve_i^\star \le C_{\alpha, \wh d} \, ,
\;
1\le i \le K \, \Big\}.
\end{equation}}
See also (\ref{b15}). To define $\wh \calB_{1-\alpha}(X)$,
we replace $\wh \calC_{1-\alpha}(X)$
by $\wh \calC_{1-\alpha}(X) \cup \calC^\star_{1-\alpha}(X)$ in
(\ref{b19}).

The resampling sample size $K$ can be specified with a leave-one-out procedure as follows.
For each $1\le i \le N$, we construct a predictive band for $Y_i(\cdot)$ conditionally on
$X_i(\cdot)$ in a similar way as above, i.e.  we construct $\wh
\calC_{1-\alpha}(X_i)$ and $\calC^\star_{1-\alpha}(X_i)$
as in (\ref{b15}) and (\ref{b18}) respectively, replacing $\wh
\eta_{j}$ by $\wh \eta_{ij} $. Leave out the term $(\wh \ve_{i1},
\cdots, \wh \ve_{i \wh d})$ in (\ref{b15}), and take the resampling sample from the other $( N -1)$ 
residual vectors. We calculate the relative frequency for the occurrence of the
event that the resulting  envelop contains $Y_i(\cdot)$
for $i=1, \cdots, N$. The hyperparameter $K$ is selected such that the corresponding relative frequency is closest to
the nominal level $1-\alpha$. For computational efficiency, one can choose $K$ among a finite
set, for example, 0, 200, 400, 600, 800 and 1000, where $K=0$ corresponds to empirical distribution without resampling.
In the cross validation approach, we use $\wh \mu_y(\cdot), \, 
\wh b_{j\ell}, \, \wh \eta_{ij}$ and $\wh \varphi_j(\cdot)$ estimated from the
whole sample of the  available observations. The leave-one-out strategy
only applies to the residuals.
Similarly, one can also develop a resampling procedure to determine the number of random curves 
to be included in the $\chi^2$-based predictive set $\wh \calC_{1-\alpha}(X)$ such that
the resulting $\wh \calB_{1-\alpha}(X)$ has the coverage probability $1 - \alpha$; see
(\ref{b20}).

\subsection{Predictive quantiles for $Y(\cdot)$ given $X(\cdot)$}

For forecasting a univariate random variable, it is informative to look at 
predictive (i.e. conditional) quantiles at different levels to gauge the
associated risk and uncertainty. 
This is 
equivalent to looking at how a predictive
interval varies  with respect to its coverage probability.
Unfortunately this analogue is no longer available in forecasting a
random vector or a random curve, for which the
concept of quantiles is not well-defined.
Nevertheless, it remains attractive to look at some ``typical"
scenarios among the curves in $\wh \calC_{1 - \alpha}(X)$ and
to observe how they vary with respect to the values of $\alpha$.

One plausible way 
is to define the most ``outside" curve (or curves) in $\wh
\calC_{1-\alpha}(X)$ as the $(1-\alpha)$-th quantile(s). 
In the context of curves, the concept of ``outsideness'' needs to be redefined as it is unlikely that
one single curve would lie completely on the one side of all the other curves in $\wh
\calC_{1-\alpha}(X)$, as the curves often cross over with each other.
\cite{narisetty2016extremal} introduced the so-called extremal depth  to
quantify the degree of ``outsideness'' for each curve. We adopt the concept of
the extremal depth to define predictive quantiles in the curve regression framework.

Let $\calS =\{ f_1, \cdots, f_p\}$ be a bundle of curves defined on set $\calI$. 
For any curve $g$ defined on $\calI$ and $u\in \calI$, a pointwise depth of $g(u)$
with respect to $\calS$ is defined as 
\[
D_g(u, \calS) = 1 - {1 \over p} \Big| \sum_{i=1}^p \big[
I \{f_i (u) < g(u) \} - I \{ f_i(u) > g(u) \} \big] \Big|,
\]
where $I(\cdot)$ denotes the indicator function.
Obviously, $D_g$ takes values $0, 1/p, 2/p, \cdots, 1$, and the larger $D_g$ is,
the more central $g(u)$ is with respect to $\{ f_i(u) \}_{i=1}^p$. Note that $g$ may or may not
be a member of $\calS$. A depth cumulative distribution function (d-CDF)
is defined as
\[
\Phi_g(r) = {1 \over A(\calI)} \int_{\calI} I\{ D_g(u, \calS) \le r \} du, \quad
r \in [0,1],
\]
where $A(\calI) = \int_{\calI}  du$.
Note that if $\Phi_g$ has most of its mass close to 0 (or 1), $g$ is away from
(or close to) the ``center'' of $\calS$. 

\cite{narisetty2016extremal} adopts the left-tail stochastic ordering for
d-CDFs to rank the ``outsideness'' or ``extremeness'' of two curves, which
differs from the other definitions for curve data depth in literature 
\citep{lopez2009concept,fraiman2001trimmed}: 
\begin{quote}
Let $0 \le u_1 \le \cdots \le u_m \le 1$ be $m$ pre-specified points.
For two curves $g$ and $h$ defined on $\calI$,  $g$ is said to be more
extremal than $h$, denoted by $g \prec h$, if there exists $1 \le k\le m$ for which
\[
\Phi_g(u_k) > \Phi_h(u_k), \quad {\rm and} \quad  \Phi_g(u_i) = \Phi_h(u_i) \;\; {\rm for
}\;\; 1\le i<k.
\]
\end{quote}
Now the extremal depth (ED) of function $g$ in relation to curve bundle $\calS$ is defined as
\[
{\rm ED}(g, \calS) = \big|\{i: f_i \preceq g, \; 1\le i \le p \} \big| \big/ p,
\]
where $| \cdot |$ denotes the cardinality of a set, and $f_i \preceq g$ if either
$f_i \prec g$ or $f_i(u_j) =g(u_j)$ for all $j=1, \cdots, m$.
Note that the smaller ${\rm ED}(g, \calS)$ is, the more outside $g$ is from $\calS$. 
We refer readers to \cite{narisetty2016extremal} for further elaboration of the ED concept.

Now we are ready to define the predictive quantile curves of $Y(\cdot)$ given
$X(\cdot)$.

\noindent
{\bf Definition 1}. For any $\alpha \in (0, 1)$, a curve $g \in \calC_{1-\alpha}(X)$
is called the $(1-\alpha)$-th predictive quantile curve of $Y(\cdot)$ given $X(\cdot)$ if
\begin{equation} \label{d4}
{\rm ED} \{ g, \calC_{1-\alpha}(X)\} = \min_{h \in \calC_{1-\alpha}(X)}
{\rm ED} \{ h, \calC_{1-\alpha}(X)\}.
\end{equation}

Intuitively, the $(1-\alpha)$-th conditional quantile is the most outside curve,
on the most possible points in $\calI_1$,
among all the curves in $\calC_{1-\alpha}(X)$.
An estimator for the quantile
curve can be obtained from replacing $\calC_{1-\alpha}(X)$ by $\wh
\calC_{1-\alpha}(X)$ in the above definition. In addition, we recommend taking 
the 2nd or even the 3rd minimizers of (\ref{d4}) as the $(1-\alpha)$-th quantiles, as 
the most outside can be either above, or below,  the other curves in $\calC_{1-\alpha}(X)$. 
Furthermore, when $C_{\alpha, \wh d}$ is estimated by the empirical distribution
of (\ref{d3}), we use the union $\wh 
\calC_{1-\alpha}(X) \cup \calC^\star_{1-\alpha}(X)$ in place of $\wh
\calC_{1-\alpha}(X)$; see (\ref{b18}).

\noindent
{\bf Remark 1}.
The definitions for $D_g(u, \calS)$ and ED$(g, \calS)$ presented
above are for $\calS$ with a finite cardinality. It serves the purpose since
in practice we always use $\wh \calC_{1-\alpha}(X)$ with finite members. 
The extension to the cases with $| \calS|=\infty$ can be formulated in terms of
distribution on $\calS$ such as
\[
D_g(u, \calS) = 1 - |E[I\{ f(u) < g(u) \} - I\{ f(u) > g(u) \}]|, \qquad
{\rm ED}(g, \calS) = P( f \preceq g),
\]
where both the expectation and the probability are taken with respect to the
distribution of $f$ on $\calS$.

\section{Simulation} \label{sec:simulation}

We illustrate the proposed PPC by simulation. 
In line with electricity load forecasting, we set $X_t(\cdot) = Y_{t-1}(\cdot)$,
and $Y_t(\cdot)$ follows the functional autoregressive (FAR) model
\begin{equation} \label{c1}
Y_t(u) = \sum_{j=1}^{d} \xi_{tj} \varphi_j(u),
\quad 
\xi_{tj} =  \sum_{\ell\in \pi_j}b_{j\ell} \xi_{t-1,\ell} + \ve_{tj}, 
\quad \ve_{tj} \sim \IID(0, \sigma^2),
\end{equation}
where $\varphi_1 (u) = 1/\sqrt{2}$ and
$\varphi_j(u) = \cos((j-1) \pi u)$ for $j\ge 2$, $u \in [-1.1]$.  We sample $d$ characteristic roots for an AR model from the interval $[2, 5]\cup[-5, -2]$ uniformly, based on which the parameters {$b_{j\ell}, \ell = 1, \cdots , d$} are determined for each $j = 1, \cdots , d$.

\noindent\textbf{Experiment 1}: Let $d=4$, $\pi_j = \{ 1, \cdots, 4\}$, $\ve_{tj}$ be normal distributed, and $\sigma=0.25$ or $0.50$
in (\ref{c1}). We draw $Y_1(\cdot), \cdots, Y_{N+N_s}(\cdot)$ from the model, and use
the first $N$ curves for estimation and the last $N_s$ curve to evaluate the
forecasting accuracy. We set the nominal coverage probability at $1- \alpha =0.9$.
For $N=100, 200, 400, 800$ or $1,600$, and $N_s =200$, we calculate the predictive set
$\wh \calC_{0.9}(\cdot)$ and 
the predictive band $\wh \calB_{0.9}(\cdot)$ based on three
methods: $\chi^2$-approximation, 
empirical distribution for
residuals (ECDF), and  ECDF with resampling (ECDF-R).
With the $\chi^2$-approximation, we include 1,500 curves in $\wh \calC_{0.9}(\cdot)$. 
For ECDF-R, we choose the resampling sample size as a multiple of 200 between 0 and 1,000.
We check the coverage rates of $\wh \calB_{0.9}(\cdot)$ for the 200 post-sample curves and draw illustrative predictive quantile curves at the $40\%$ and $90\%$ confidence levels. 
For more comprehensive evaluation,  we also
calculate the pointwise mean absolute errors (MAE) on a 
grid $u_i = -1 + 0.04(i-1)$ $(i=1, \cdots, 51)$ over the 200 post-sample curves $\{ Y_k(\cdot)\}$:
\begin{equation} \label{c2}
{\rm \bf{MAE}} = {1 \over 51 \times 200 } \sum_{k=1}^{200}
\sum_{i=1}^{51} | \wh Y_k(u_i) - Y_k(u_i) |.
\end{equation}
For each setting, we replicate the above exercise 400 times.

\begin{figure}[!t]
	\centering
	\begin{subfigure}{0.85\textwidth}
		\centering
		\includegraphics[width=1\linewidth]{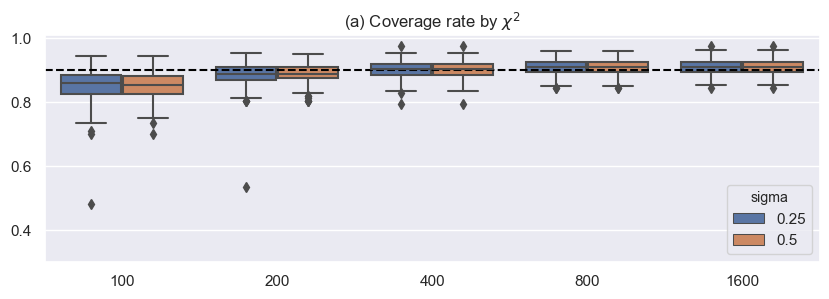}
	\end{subfigure}
	\begin{subfigure}{.85\textwidth}
		\centering
		\includegraphics[width=1\linewidth]{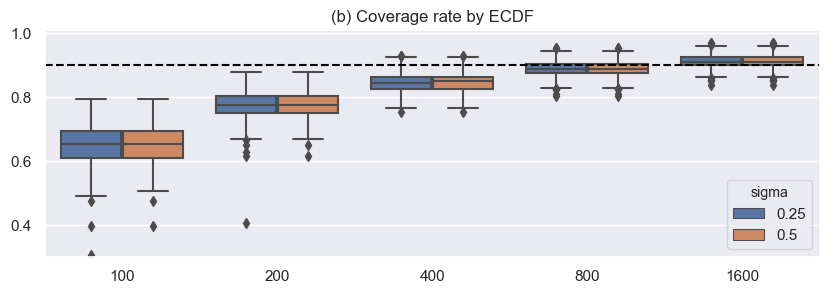}
	\end{subfigure}
	\begin{subfigure}{.85\textwidth}
		\centering
		\includegraphics[width=1\linewidth]{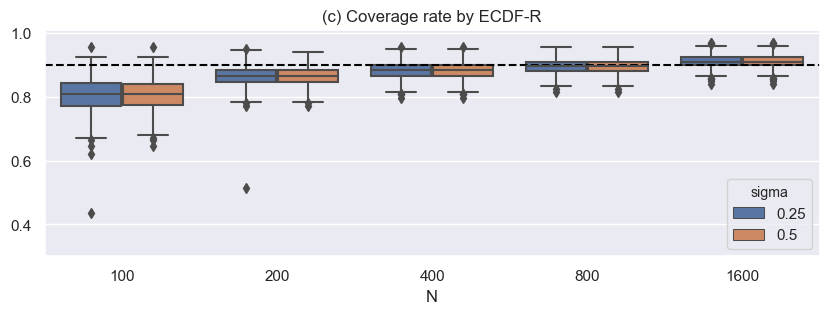}
	\end{subfigure}
	\caption{\small Experiment 1 --  Box plots of post sample coverage rates
		of  $\wh \calB_{0.9}(X)$ based on (a) $\chi^2$, (b) ECDF, 
		and (c) ECDF-R for $N=100, 200, 400, 800, 1600$
		and $\sigma =0.25, 0.5$. The horizontal dash lines mark the positions of the nominal coverage probability 0.9.} 
	\label{fig:1}
\end{figure}

Figure~\ref{fig:1} displays the box plots of the coverage rates of
$\wh \calB_{0.9}(\cdot)$ based on, $\chi^2$-approximation,
ECDF and ECDF-R respectively, for the 200 post-sample curves over 400
replications. Since $\ve_{tj}$ are normal distributed, $\chi^2$-based $\wh \calB_{0.9}(\cdot)$
performs the best. The sample size has minimal impact for predictive region with $\chi^2$, where the coverage rates are all close to $0.9$ except for $N=100$.  A simple trial shows that, when using $10,000$ instead of $1,500$ simulated curves, the coverage rate can be corrected to be around the nominal rate for $N = 100$. It is also clear that the resampling adjustment for
the ECDF based method is necessary, as the
coverage rates of $\wh \calB_{0.9}(\cdot)$ based on ECDF-R
are clearly closer to the nominal coverage probability 0.9 than those
based on ECDF,
especially for $N\le 400$.
Also clearly noticeable is the improvement of performance as the sample size
$N$ increases. On the other hand, the different noise levels $\sigma =0.25 $ and 0.50
has no impact on the performance, as the signal-to-noise ratio of a stationary
AR process is invariant with respect to the noise level.

\begin{table}[!b]
	\centering
	\caption{\small Experiment 1 -- The means and standard errors (in parentheses)
		of MAE in (\ref{c2}), the average length (AvL) of $\wh
		\calB_{0.9}(\cdot)$, the proportion of the overlapping area (POA) in (\ref{c3}),
		and $\wh d$ over 400 replications.
		For $\wh \calB_{0.9}(\cdot)$ based on ECDF-R, the means and standard errors 
		of the selected resampling sample size $K$ are also included.}
	\makebox[\textwidth]{\resizebox{1\linewidth}{!}{
			\begin{tabular}{r|r|r|r|rr|rr|rrr}
				\hline
				\hline
				&       &       &       & \multicolumn{2}{c|}{\textbf{$\chi^2$}} & \multicolumn{2}{c|}{\textbf{ECDF}} & \multicolumn{3}{c}{\textbf{ECDF-R}} \\
				\cline{5-11}  
				\multicolumn{1}{c|}{\textbf{$\sigma$}} & \multicolumn{1}{c|}{\textbf{$N$}} &  \multicolumn{1}{c|}{\textbf{$\hat{d}$}} & \multicolumn{1}{c|}{\textbf{MAE}} &  \multicolumn{1}{c}{\textbf{AvL}} & \multicolumn{1}{c|}{\textbf{POA}} &  \multicolumn{1}{c}{\textbf{AvL}} & \multicolumn{1}{c|}{\textbf{POA}} &  \multicolumn{1}{c}{\textbf{AvL}} & \multicolumn{1}{c}{\textbf{$K$}} & \multicolumn{1}{c}{\textbf{POA}} \\
				\hline
				\textbf{0.25} & \textbf{100} & 4     & .297(.035) &  1.726(.063) & .948(.028) &  1.420(.065) & .829(.038) &  1.626(.075) & 802(190) & .919(.034) \\
				& \textbf{200} & 4        & .288(.016) &  1.753(.045) & .970(.015)  & 1.549(.054) & .900(.026) &  1.690(.056) & 779(192) & .954(.020) \\
				& \textbf{400} & 4      & .284(.008) &  1.765(.036) & .980(.009)  & 1.648(.044) & .947(.017) &  1.722(.045) & 639(220) & .971(.013) \\
				& \textbf{800} & 4       & .282(.008) &  1.771(.029) & .985(.007)  & 1.723(.034) & .975(.010) &  1.736(.033) & 173(184) & .978(.010) \\
				& \textbf{1600} & 4         & .282(.008) &  1.777(.024) & .988(.006)  & 1.779(.024) & .989(.006) &  1.779(.024) & 0(0) & .989(.006) \\
				\hline
				\textbf{0.50} & \textbf{100} & 4     & .595(.098) & 3.447(.122) & .949(.030) &  2.835(.128) & .829(.039) &  3.250(.149) & 814(184) & .920(.036) \\
				& \textbf{200} & 4      & .575(.016)  & 3.506(.087) & .970(.012) &  3.097(.103) & .901(.024) &  3.379(.109) & 783(194) & .954(.018) \\
				& \textbf{400} & 4         & .567(.016)  & 3.530(.071) & .980(.008) &  3.296(.087) & .947(.016) &  3.443(.090) & 637(222) & .971(.012) \\
				& \textbf{800} & 4        & .564(.016)  & 3.542(.057) & .985(.007) &  3.444(.068) & .975(.010) &  3.471(.066) & 174(185) & .978(.010) \\
				& \textbf{1600} & 4        & .563(.015) & 3.553(.048) & .988(.006) &  3.558(.049) & .989(.006) &  3.558(.049) & 0(0) & .989(.006) \\
				\hline
				\hline
			\end{tabular}%
	}}
	\label{tab:1}%
\end{table}

Table~\ref{tab:1} lists the means and standard errors (in parentheses),
over 400 replications,
of MAE in (\ref{c2}), the average length (AvL) of $\wh
\calB_{0.9}(\cdot)$ over the 51
grid points, and the proportion of the overlapping area:
\begin{equation} \label{c3}
{\rm POA} =
{\rm Area}\{ \wh
\calB_{0.9}(\cdot) \cap \calB_{0.9}(\cdot)\} \big/ {\rm Area}\{ \calB_{0.9}(\cdot)\},
\end{equation}
where Area$\{ \calB_{0.9}(\cdot)\}$ is evaluated by simulation from the
true model. 
The AvL for $\wh\calB_{0.9}(\cdot)$ based on $\chi^2$ or ECDF-R tends to
be larger than that based on ECDF. This is due to the fact that $\wh\calB_{0.9}(\cdot)$
based on ECDF tends to be smaller, reflected by lower coverage rates (see
Figure \ref{fig:1}) and smaller overlapping areas.
Since the overlapping areas are always at least, respectively, 94.8\%,
82.9\%, 91.9\% based on $\chi^2$, ECDF, ECDF-R, $\wh\calB_{0.9}(\cdot)$
provides an accurate estimator for $\calB_{0.9}(\cdot)$.
The average length of $\wh\calB_{0.9}(\cdot)$ with $\sigma=0.50$ is significant
larger than that with $\sigma=0.25$, reflecting more uncertainty in forecasting
due to large noise.
Note that the true AvL is 1.66 with $\sigma = 0.25$, and 3.32 with $\sigma = 0.50$.
The improvement due to the increase of $N$ is evident.
For $\wh \calB_{0.9}(\cdot)$ based on ECDF-R, we also report the mean and
standard errors for $K$. The selected value of $K$ drops when sample size $N$ increases, indicating the decreasing need of resampling adjustment for large $N$. For $N=1,600$, the choice is direct construction based on the empirical distribution, i.e. $K=0.$ 
Also included in the table are the average estimated values for $\wh d$. 
In fact, $\wh d$ is always equal to the true value 4 in the 400 replications.

\begin{figure}[!b]
	\begin{center}
		\begin{subfigure}{0.49\textwidth}
			\centering
			\includegraphics[width=1\linewidth]{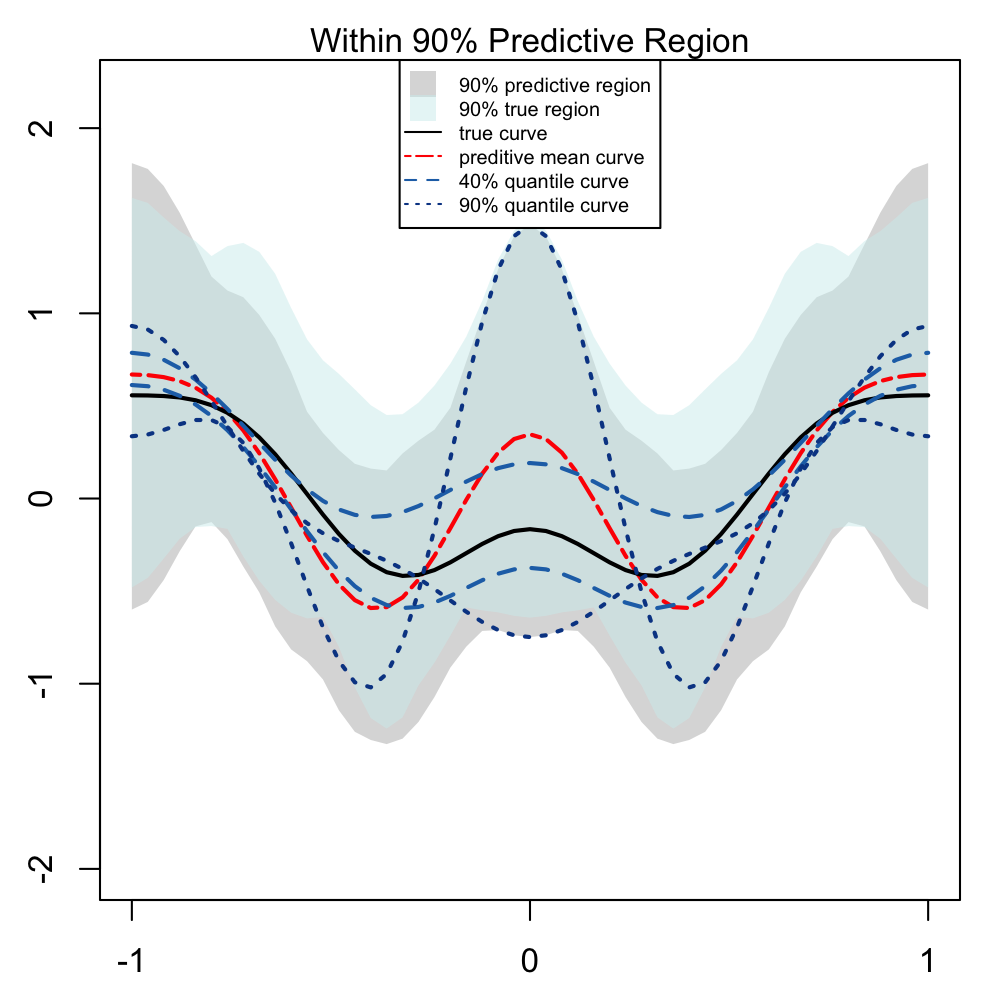}
		\end{subfigure}
		\begin{subfigure}{.49\textwidth}
			\centering
			\includegraphics[width=1\linewidth]{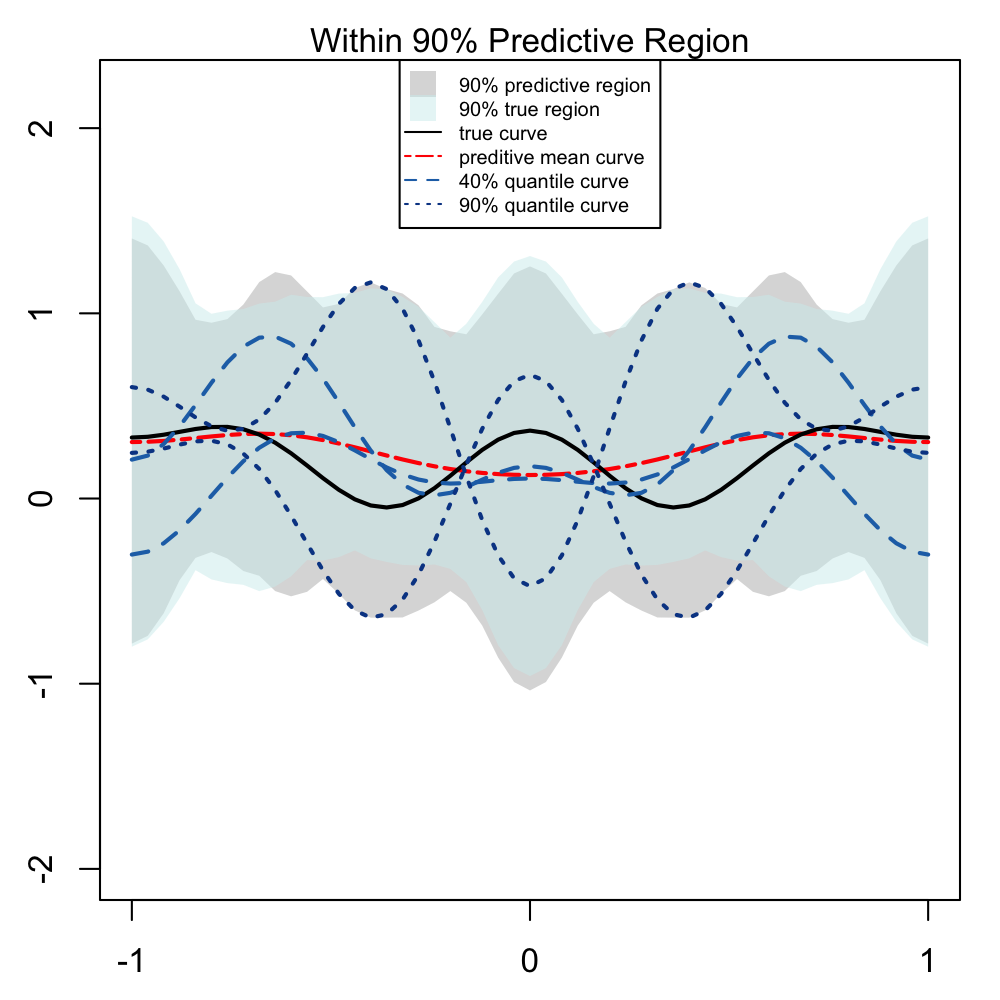}
		\end{subfigure}
	\end{center}
	\caption{\small Experiment 1-- Randomly selected 2 post-sample curves (solid lines), their true 90\% regions (light blue shadow areas), predictive mean curves (dash-dotted lines), 90\% predictive regions (grey shadow areas) and  quantile curves at 90\% and 40\% confidence levels (dotted and dashed lines respectively) based on  $\chi^2$, for $N=100$ and $\sigma=0.25$.} 
	\label{fig:3} 
\end{figure}

As illustration, Figure~\ref{fig:3} depicts $2$ randomly selected
post-sample curves together with their 90\% predictive regions and
quantile curves at 40\% and 90\% confidence levels based on $\chi^2$ 
with $N=100$ and $\sigma = 0.25$. Even for the small sample size case,
the 90\% predictive regions of the 2  curves are overlapping with the
true regions  to a large extent. In addition,  the 90\% predictive
quantile curves derived from ED coincide with some parts of the bounds of
90\% predictive regions, and indeed represent the most ``outside'' curve
in the associated predictive regions. The predictive quantile curves at a
higher confidence level (90\%) are more ``outside'' than those at a lower
level (40\%) for most parts of the domain of the curves.

\askip

\noindent
{\bf Experiment 2}: We set in (\ref{c1}) $d=6$, $\pi_j= \{ 1, \cdots, 6\}$, and consider three
distributions for $\ve_{tj}$: $N(0, \sigma^2)$, the centered and rescaled standard
 exponential distribution, and the rescaled $t$-distribution with 5 degrees of
freedom. The rescaling makes the standard deviation $\sigma =0.25$ for
all the three distributions.
 Since  PPC based on ECDF-R are clearly superior than
those based on ECDF in Experiment 1, we drop the results based on ECDF. 

\begin{table}[tb]
	\caption{\small Experiment 2 -- The means and standard errors (in parentheses) of
		MAE in (\ref{c2}), the coverage rate (CR) and the average length (AvL) of
		$\wh \calB_{0.9}(\cdot)$, the proportion of the overlapping area (POA) in (\ref{c3}),
		and $\wh d$ over 400 replications. For $\wh \calB_{0.9}(\cdot)$ based on ECDF-R, the means and standard errors 
		of the selected resampling sample size $K$ are also included.}
	\label{tab:2}%
	\makebox[\textwidth]{\resizebox{1\linewidth}{!}{
			\begin{tabular}{r|c|r|r|rrr|rrrr}
				\hline
				\hline
				&       &       &            & \multicolumn{3}{c|}{\textbf{$\chi^2$}} & \multicolumn{4}{c}{\textbf{ECDF-R}} \\
				\cline{5-11}    \multicolumn{1}{c|}{\textbf{$N$}} & \textbf{dist} &
				\multicolumn{1}{c|}{\textbf{$\hat{d}$}} &
				\multicolumn{1}{c|}{\textbf{MAE}} & \multicolumn{1}{c}{\textbf{CR}} &
				\multicolumn{1}{c}{\textbf{AvL}} &
				\multicolumn{1}{c|}{\textbf{POA}} &
				\multicolumn{1}{c}{\textbf{CR}} & \multicolumn{1}{c}{\textbf{AvL}} &
				\multicolumn{1}{c}{\textbf{$K$}} &
				\multicolumn{1}{c}{\textbf{POA}} \\
				\hline
				\textbf{100} & \textbf{norm} & 6     & .364(.010) & .849(.040) & 2.298(.069) & .952(.016) &  .787(.053) & 2.138(.081) & 841(172) & .920(.024)  \\
				& \textbf{t}$_5$ & 6        & .355(.014) & .851(.045) & 2.288(.133) & .952(.022) &  .806(.055) & 2.146(.128) & 818(188) & .926(.032)  \\
				& \textbf{exp} & 6         & .349(.015) & .850(.040) & 2.284(.128) & .951(.025)   & .820(.052) & 2.171(.151) & 798(187) & .931(.033) \\
				\hline
				\textbf{200} & \textbf{norm}   & 6     & .353(.009) & .890(.029) & 2.332(.054) & .969(.011) &  .859(.039) & 2.237(.073) & 811(171)  & .953(.017) \\
				& \textbf{t}$_5$ & 6          & .344(.011) & .886(.031) & 2.325(.092) & .972(.014) & .866(.034) & 2.251(.093) & 810(186)& .960(.019)  \\
				& \textbf{exp} & 6       & .339(.012) & .879(.031) & 2.324(.106) & .969(.018) & .870(.035) & 2.289(.130) &  797(193)& .963(.022)  \\
				\hline
				\textbf{400} & \textbf{norm} & 6        & .349(.008) & .907(.023) & 2.347(.042) & .978(.009) & .886(.025) & 2.278(.052) & 645(201) & .968(.012)  \\
				& \textbf{t}$_5$ & 6          & .340(.010) & .902(.024) & 2.346(.070) & .980(.012) & .888(.028) & 2.301(.075) & 672(202) & .972(.014) \\
				& \textbf{exp} & 6       & .333(.011) & .892(.025) & 2.344(.079) & .977(.014) &  .891(.025) & 2.337(.094) & 656(190) & .976(.016) \\
				\hline
				\textbf{800} & \textbf{norm} & 6    & .348(.007) & .911(.022) & 2.352(.036) & .983(.007) &  .890(.027) & 2.291(.040) & 163(137) & .974(.010) \\
				& \textbf{t}$_5$ & 6       & .338(.009) & .905(.023) & 2.355(.053) & .983(.011) &  .893(.024) & 2.313(.055) & 347(145) & .976(.014)  \\
				& \textbf{exp} & 6      & .332(.013) & .895(.025) & 2.356(.057) & .981(.012) &  .893(.028) & 2.332(.071) & 392(169) & .980(.013)  \\
				\hline
				\textbf{1600} & \textbf{norm} & 6    & .346(.008) & .916(.022) & 2.359(.032) & .985(.007) &  .918(.022) & 2.365(.034) & 0(0) & .986(.007)  \\
				& \textbf{t}$_5$ & 6      & .338(.011) & .907(.022) & 2.360(.045) & .986(.009) &  .902(.022) & 2.345(.044) & 6(38) & .983(.010)   \\
				& \textbf{exp} & 6        & .331(.012) & .898(.023) & 2.359(.042) & .983(.011) & .898(.023) & 2.318(.046) & 36(79) & .982(.011)  \\
				\hline
				\hline
			\end{tabular}%
	}}
\end{table}

We adopt the similar setting as in Experiment 1. The results are reported in
Tables~\ref{tab:2}. The performance of both the pointwise and probabilistic forecasting are satisfactory,
and are about the same with the three different distributions for $\ve_{ti}$.
The coverage rates (CR) of $\wh \calB_{0.9}(\cdot)$
are almost as good as in Experiment 1, especially for $N\ge 400$.
In terms of CR, AvL and POA, there is hardly any substantial
difference between $\wh \calB_{0.9}(\cdot)$ based on $\chi^2$ or on ECDF-R. 
It is somehow surprising that the $\chi^2$ based method works fine with
the heavy-tailed $t_5$ distribution and the highly skewed exponential distribution. 
Note that the comparison of AvL can only be made for the two cases with
about the same CR. When $N=1,600$ with exponential distributed $\ve_{ti}$, the CR for
$\wh \calB_{0.9}(\cdot)$
based on the two methods are the same, and the AvL based on ECDF-R is 2.318
which is smaller than 2.359, the AvL based on $\chi^2$. This may be due
to the fact that the residuals from the fitted model capture the skewed
exponential distribution better than the $\chi^2$ approximation, though
one may argue if such a difference is really substantial.
The true AvL is 2.17, 2.13 and 2.10, respectively, with normal, $t_5$ and
exponential distributed innovations.

\askip

\noindent
{\bf Experiment 3}: We investigate the performance of PPC for higher
order curve regressions. Consider an FAR(3) process, $Y_t(\cdot)$ is defined
as in (\ref{c1}) where the second equation is
replaced by one of the three equations below.
\begin{align*}
\mbox{Non sparse: }&
\xi_{tj} =  \sum_{l=1}^{d}b_{jl} \xi_{t-1,l} +\sum_{l=1}^{d}b_{j,l+d}
\xi_{t-2,l} +\sum_{l=1}^{d}b_{j,l+2d} \xi_{t-3,l} + \ve_{tj},\\
\mbox{Lag sparse: }&
\xi_{tj} =   \sum_{\ell\in \pi_j}b_{j\ell} \xi_{t-3,l} + \ve_{tj}, \quad
\pi_j= \{ 1, \cdots, 6\},\\
\mbox{Diagonal sparse: }&
\xi_{tj} =  b_{j1} \xi_{t-1,j} + b_{j,2} \xi_{t-2,j} + b_{j,3} \xi_{t-3,j} + \ve_{tj}.
\end{align*}
Similar to Experiment 1, we set $d=4$ and let $\ve_{tj} \sim
N(0,\sigma^2)$ with $\sigma = 0.25$, but now $X_t = [Y_{t-1},Y_{t-2},Y_{t-3}]'$.



\begin{table}[!t]
	\centering
	\caption{\small Experiment 3 -- The means and standard errors (in parentheses) of
		MAE in (\ref{c2}), the coverage rate (CR) and the average length (AvL) of
		$\wh \calB_{0.9}(\cdot)$, the proportion of the overlapping area (POA) in (\ref{c3}),
		and $\wh d$ over 400 replications. For $\wh \calB_{0.9}(\cdot)$ based on ECDF-R, the means and standard errors 
		of the selected resampling sample size $K$ are also included.}
	\label{tab:3}%
	\makebox[\textwidth]{\resizebox{1\linewidth}{!}{
			\begin{tabular}{r|r|r|r|rrr|rrrr}
				\hline
				\hline
				&       &       &       & \multicolumn{3}{c|}{\textbf{$\chi^2$}} & \multicolumn{4}{c}{\textbf{ECDF-R}} \\
				\cline{5-11}    \multicolumn{1}{l|}{\textbf{sparsity}} & \multicolumn{1}{l|}{\textbf{N}} & \multicolumn{1}{l|}{\textbf{$\wh d$}} & \multicolumn{1}{l|}{\textbf{MAE}} & \multicolumn{1}{c}{\textbf{CR}} & \multicolumn{1}{c}{\textbf{AvL}} & \multicolumn{1}{c|}{\textbf{POA}} & \multicolumn{1}{c}{\textbf{CR}} & \multicolumn{1}{c}{\textbf{AvL}} & \multicolumn{1}{c}{\textbf{$K$}} & \multicolumn{1}{c}{\textbf{POA}} \\
				\hline
				\multicolumn{1}{l|}{\textbf{non sparse}} & \textbf{100} & 4     & .310(.017) & .834(.047) & 1.768(.090) & .939(.020) & .773(.061) & 1.645(.093) & 816(199) & .905(.029) \\
				& \textbf{200} & 4     & .296(.016) & .882(.031) & 1.787(.096) & .962(.013) & .849(.037) & 1.712(.095) & 781(181) & .944(.019) \\
				& \textbf{400} & 4     & .289(.013) & .900(.026) & 1.790(.062) & .976(.010) & .881(.032) & 1.740(.061) & 595(228) & .964(.015) \\
				& \textbf{800} & 4     & .286(.010) & .905(.024) & 1.785(.043) & .980(.008) & .889(.025) & 1.744(.042) & 173(182)& .972(.011) \\
				& \textbf{1600} & 4     & .285(.012) & .910(.023) & 1.792(.059) & .986(.007) & .911(.022) & 1.795(.058) & 0(0) & .986(.007) \\
				\hline
				\multicolumn{1}{l|}{\textbf{lag sparse}} & \textbf{100} & 4     & .302(.014) & .832(.046) & 1.714(.071) & .936(.022) & .780(.058) & 1.613(.082) & 795(187) & .906(.031) \\
				& \textbf{200} & 4     & .292(.011) & .880(.030) & 1.756(.054) & .963(.013) & .851(.037) & 1.689(.062) & 787(198) & .945(.018) \\
				& \textbf{400} & 4     & .286(.009) & .898(.025) & 1.772(.043) & .976(.008) & .881(.029) & 1.730(.052) & 624(227) & .967(.013) \\
				& \textbf{800} & 4     & .284(.009) & .905(.023) & 1.779(.037) & .982(.007) & .892(.025) & 1.743(.038) & 194(184)& .975(.009) \\
				& \textbf{1600} & 4     & .284(.008) & .910(.023) & 1.785(.028) & .987(.006) & .909(.023) & 1.786(.030) & 0(0) & .987(.007) \\
				\hline
				\multicolumn{1}{l|}{\textbf{diag sparse}} & \textbf{100} & 4     & .332(.175) & .803(.104) & 1.705(.070) & .920(.062) & .750(.107) & 1.606(.079) & 742(218) & .891(.065) \\
				& \textbf{200} & 4     & .297(.033) & .864(.053) & 1.746(.050) & .955(.026) & .838(.060) & 1.685(.060) & 783(195) & .940(.030) \\
				& \textbf{400} & 4     & .290(.033) & .893(.029) & 1.777(.146) & .973(.011) & .876(.035) & 1.734(.134) & 618(211) & .964(.014) \\
				& \textbf{800} & 4     & .289(.046) & .904(.023) & 1.801(.286) & .981(.008) & .893(.026) & 1.763(.246) & 172(169) & .973(.010) \\
				& \textbf{1600} & 4     & .282(.010) & .911(.020) & 1.781(.041) & .986(.007) & .913(.021) & 1.784(.031) & 0(0) & .987(.007) \\
				\hline
			\end{tabular}%
	}}
\end{table}%

Table~\ref{tab:3}  shows the detailed predictive results based on either
$\chi^{2}$ or  ECDF-R, for the non-sparse FAR(3) and the two sparse
FAR(3) models above. The first $10$ singular value components of the
regressor curve are included for the AIC selection. The true AvL is
1.66 for all the three FAR(3) models. In general, PPC work well for the
higher order curve regressions, with CRs close to the nominal coverage
probability $0.9$, POA above $90\%$ and AvL close to the true AvL. The
patterns of CRs and POAs are very similar to those in Experiment 1,
though with slightly worse performance for small sample size ($N \leq
400$). 
Performance for the lag sparse model is better than the non-sparse
one, producing narrower AvLs for similar coverage rates. When comparing
the diagonal sparse model with non-sparse model, there is no clear pattern.
Note that our estimation method makes no use of the information of
the particular sparse structures.


\section{Probabilistic forecasting for daily electricity loads} \label{sec:realdata}

In this section, we apply the proposed PPC to a real data set consisting of
French daily electricity load curves from January 1, 2012 to December 31, 2019.

%

\subsection{Data}

The French electricity consumption data are collected from the website of the system operator RTE (R\'{e}seau et Transport d'\'{E}lectricit\'{e}): \url{https://opendata.rte-france.com}) at a temporal resolution of every half-hour (i.e. 48 points on each day). We remove the
data on public holidays, the day before and the day after the holidays, and
also in the Christmas periods. Empirical experience indicates that the electricity
demand on those days behaves differently, and requires different treatment.

Temperature  is a key exogenous factor for electricity consumption in 
France  due to electrical heating and cooling. We obtained data from 96
meteostations in France from the website of the French weather forecaster
M\'{e}t\'{e}oFrance (\url{https://donneespubliques.meteofrance.fr/}).
Temperature data are provided at a three hours resolution and
interpolated with natural cubic splines at a half-hour resolution. Figure
\ref{fig:eda} displays the dynamic evolution of the daily load curves
from 2012 to 2019, the corresponding daily temperature curves and a
scatter plot showing the
strong dependence between the load and the temperature.


\begin{figure}[!t]
	\centering
		\begin{subfigure}{0.9\textwidth}
			\centering
			\includegraphics[width=1\linewidth]{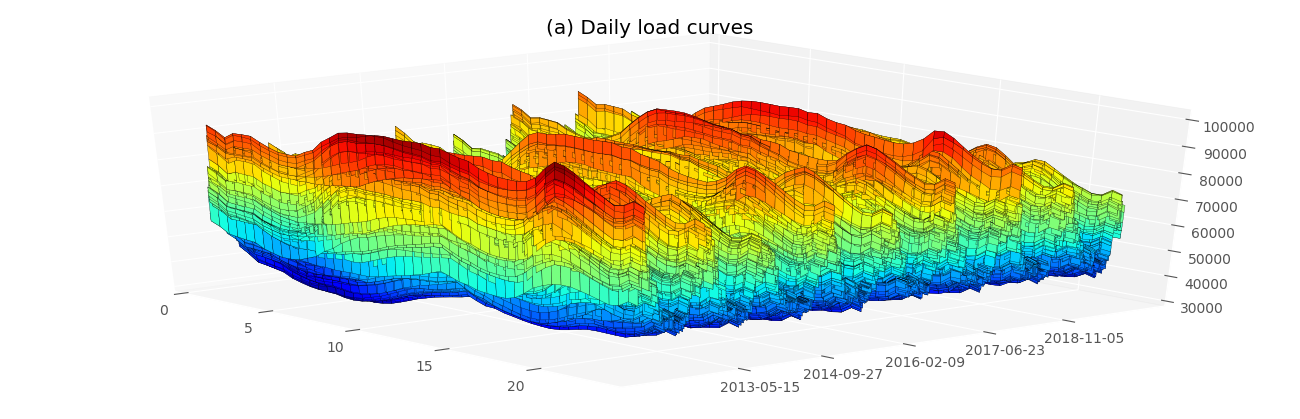}
		\end{subfigure}
		\begin{subfigure}{.9\textwidth}
			\centering
			\includegraphics[width=1\linewidth]{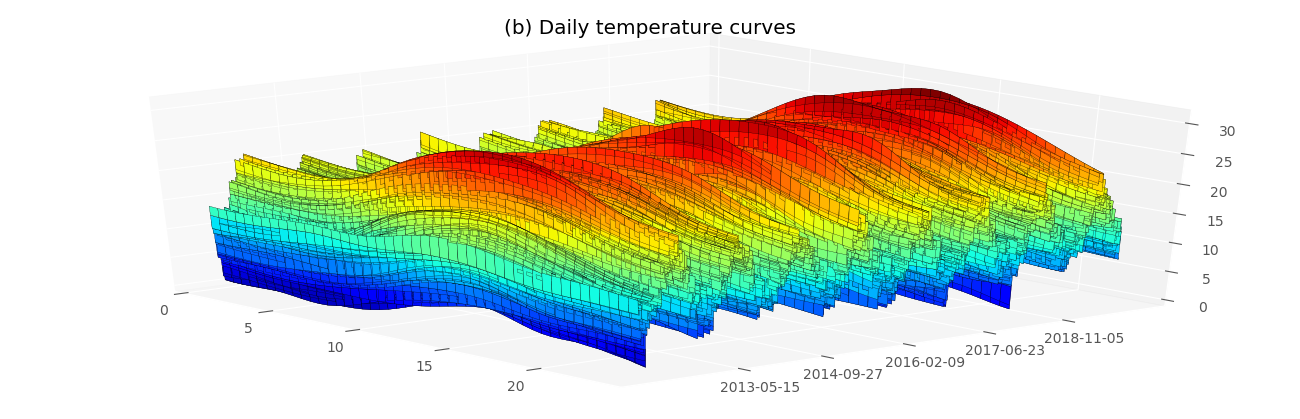}
		\end{subfigure}
		\begin{subfigure}{.80\textwidth}
		\centering
		\includegraphics[width=1\linewidth]{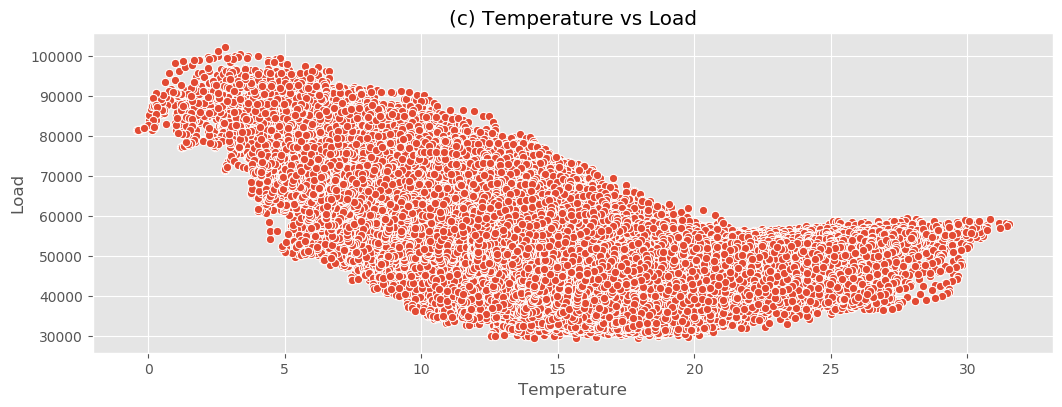}
	\end{subfigure}
	\caption{\small (a) Daily loads curves (b) Daily
temperature curves (c) Scatter plot of temperature vs
loads at half-hour frequency. Data are from 2012 to 2019.}
	\label{fig:eda}
\end{figure}

Electricity loads, like other energy data, highly depend on economic and
meteorological factors. The changes of temperature  introduce seasonality
in the demand which is higher in winter and lower in
summer. The shift of working routines causes varying diurnal patterns
between weekdays and weekends. Therefore, the profiles of daily load
curves differ in months and days. As an illustration, Figure
\ref{fig:nonstationarity} depicts the daily curves on
Tuesday-Wednesday in June, 
Saturday-Sunday in June and Tuesday-Wednesday in November 
between 2012 to 2018. In June, the curves on
Tuesday-Wednesday are similar, but differ from those on
Saturday-Sunday. Furthermore, the demand is higher 
on weekdays than that in weekends. In addition, the diurnal pattern varies
over different months. For example, peaks occur at noon in June (due to 
cooling consumption) while in the evening in November (due to 
heating consumption).
The inhomogeneous phenomena dictates
the need to segment days into different homogeneous
groups, as the proposed PPC are developed under
a stationary framework. 
Table \ref{tab:subgroups} lists the segmentation rule adopted by EDF:
each week is divided into 5 groups and the 12 months are classified into
7 groups, leading to in total 35 groups. We will fit a separate model
for each of the 35 groups.
\begin{figure}[!t]
	\centering
	\includegraphics[width=0.9\linewidth]{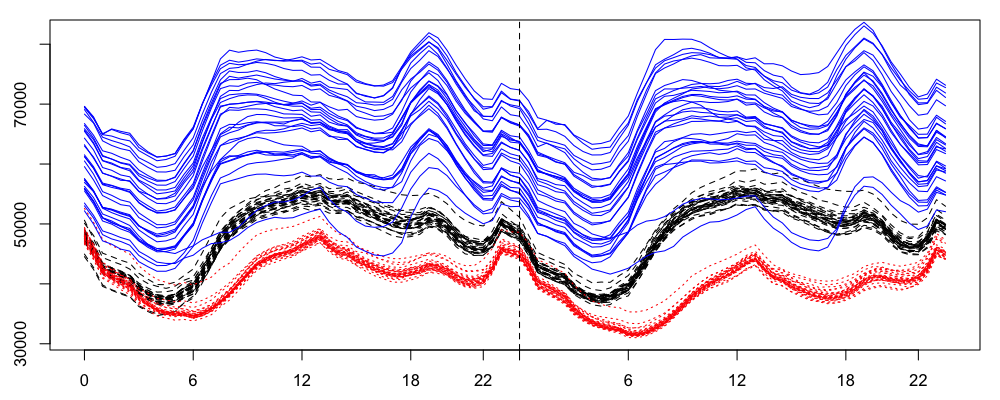}
	\caption{\small Electricity load curves on Tuesday-Wednesday in
June (black dashed lines), Saturday-Sunday in June (red dotted lines) and
Tuesday-Wednesday in November (blue solid lines) between 2012 to 2018.}
	\label{fig:nonstationarity}
\end{figure}


\begin{table}[!t]
  \centering
  \caption{Classification rule for electricity load data}
   \makebox[\textwidth]{\resizebox{1\linewidth}{!}{
    \begin{tabular}{c|c|c|c|c|c|c|c}
\hline\hline  \textbf{Day Class} & \textbf{1}     & \textbf{2}     & \textbf{3}     & \textbf{4}     & \textbf{5}     \\
\cline{1-6}    \textbf{Day type} & Mon   & Tue, Wed, Thu & Fri   & Sat   & Sun    \\
    \hline
    \textbf{Month Class} & \textbf{1}     & \textbf{2}     & \textbf{3}     & \textbf{4}     & \textbf{5}     & \textbf{6}     & \textbf{7} \\
    \hline
    \textbf{Month} & Dec, Jan, Feb & Mar   & Apr, May & Jun, Jul, Sep & Aug & Oct   & Nov \\
    \hline\hline
    \end{tabular}%
}}
  \label{tab:subgroups}%
\end{table}%


\subsection{Probabilistic forecasting}

For each of the 35 groups, we pair the daily load curve $Y_t(\cdot)$ on day $t$  and $X_t(\cdot)$ consisting of three curves: the load curve on the previous day $Y_{t-1}(\cdot)$, the load curve one week earlier $Y_{t-7}(\cdot)$, and the daily temperature curve $T_t(\cdot)$.  The temperature forecasts can be obtained from e.g. M\'et\'eo-France, which maintain a stable high accuracy. We thus directly use the actual temperature in our analysis, which is commonly adopted in the literature of electricity load forecasting.
To make the three curves on the
same scale, we standardize each of them (within each of the 35 groups)
first before combining them into $X_t(\cdot)$. The standardization uses
the training data only, see below.

We use the data in 2019 to evaluate the post-sample forecasting performance.
More precisely, for each day in 2019, we use all the data in the same 
group before that day to fit the model. The day-ahead forecasts are produced at noon of each day, in alignment with the forecasting operation in EDF.  The sample size $N$ varies among the 35 groups
from  $24$ to $284$. For determining $\wh \pi_j$ in (\ref{equ:1}) by AIC,
we  include the first $\min(\lceil N/2\rceil,48)$ $\wh \eta_{tj}$ as the candidate regressors to avoid over-fitting problem for small $N$.
We set the nominal coverage probability at $1-\alpha =0.9$. The predictive set $\wh \calC_{0.9}(\cdot)$ based on $\chi^2$-approximation
contains $K_{\chi^2}$ randomly generated curves, where $K_{\chi^2} = 5,000$ or 20,000. For $\wh \calC_{0.9}(\cdot)$ based on ECDF-R,  we choose the resampling sample size $K$ as a multiple of 800 between 0 and 4,800 using the proposed leave-one-out procedure.

For the comparison purpose, we also include three state-of-art models popular in electricity load forecasting. Those models are designed for forecasting the load
at time point $h$ on day $t$, denoted by $y_{t,h}$. Fitting these models separately for each  hour of the day is a common practice in electricity load forecasting as hour of the day plays a prominent role.

\noindent
1. The generalized additive model (GAM):
\begin{align*}
y_{t,h} = & \psi_{D_{t},h} + f_{1,h}(t) + f_{2,h}(S_t)+ f_{3,h}(y_{t-1,h}, D_t) + f_{4,h}(y_{t-7,h}) + f_{5,h}(t,T_{t,h}) \\
& + f_{6,h}(T_{t,h}^s(0.95))  + f_{7,h}(T_{t,h}^s(0.99)) + f_{8,h}(T_{t,h}^{\max},T_{t,h}^{\min}) + \epsilon_{t,h},
\end{align*}
where $T_{t,h}$ denotes the temperature at time $h$ on day $t$, $T_{t,h}^{\max},
T_{t,h}^{\min}$ are the daily maximum, minimum temperatures, $T_{t,h}^s(v)  =
v T_{t-1,h}^s(v) + (1-v)T_{t,h}$ is an exponentially smoothed version of $T_{t,h}$,
 $S_t$ represents an annual cycling effect,  $D_t$ denotes the day effect and each $f_{j,h}$ is a smooth function of the covariates with the thin plate regression splines basis functions.
 
\noindent
2. The seasonal autoregressive (SAR) model: 
\[
y_{t,h}-\mu_{y,h} = \omega_{1,h}\left(y_{t-1,h}-\mu_{y,h}\right)+\omega_{2,h}\left(y_{t-2,h}-\mu_{y,h}\right) +\omega_{3,h}\left(y_{t-7,h}-\mu_{y,h}\right) + \epsilon_{t,h}.
\]

\noindent
3. The SAR with exogenous variable (SARX) model:
\begin{align*}
y_{t,h}-\mu_{y,h} = & \omega_{1,h}\left(y_{t-1,h}-\mu_{y,h}\right)+\omega_{2,h}\left(y_{t-2,h}-\mu_{y,h}\right) +\omega_{3,h}\left(y_{t-7,h}-\mu_{y,h}\right)\\
& + \omega_{T,h}(T_{t,h}-\mu_{T,h})  + \epsilon_{t,h}.
\end{align*}

The GAM was proposed by \cite{pierrot2011short} and engineered by EDF. It serves here as an industry benchmark.
The SAR and SARX are implemented periodically for series attached to each half-hour and fitted separately for each of the 35 groups. 
For a fair comparison with our curve regression approach, all the forecasts for the
next 48 points are made at noon, with forecast time horizons $h=1,  \cdots, 48$
respectively. More precisely, 
an ``error correction'' is applied as follows to each of GAM, SAR and SARX
to incorporate the intraday dependence:
we calculate the 10-fold cross validation predictive errors (PE) for
each of the 48 half-hour models, 
and combine them together to form a half-hourly PE series.
We fit an ARMA model to this combined PE series, which is then
used to predict the 48 future PE at time horizons $1 , \cdots, 48$ respectively.
The pointwise predictor for $y_{t,h}$ is defined as
 $\wh y_{t,h} = \wt y_{t,h}+ \wh m$,
and the predictive interval for $y_{t,h}$ is
$[\wt y_{t,h}+ \wh Q_1, \wt y_{t,h} +\wh Q_2]$,
 where $\wt y_{t,h}$ is the point predictor based on
GAM, SAR or SARX, and $\wh m$ and $[\wh Q_1, \wh Q_2]$  denote, respectively,
 the point predictor and the interval predictor for the corresponding
 PE from the ARMA model.

\begin{table}[!t]
  \centering
  \caption{\small The mean absolute percentage error (MAPE), coverage
rate (CR), pointwise coverage rate (PCR), and the average length  (AvL)
of the predictive intervals
at the 48 half-hour points of the different forecasting methods.
The nominal coverage probability is 0.9.}
    \begin{tabular}{l|c|c|c|c|c}
    \hline
    \hline
    \multicolumn{1}{p{6em}|}{Method} &
    \multicolumn{1}{p{2em}|}{$\hat{d}$} & \multicolumn{1}{p{4em}|}{MAPE} & 
CR & PCR& AvL\\
    \hline
    $\wh \calB_{0.9}(\cdot)$ based on $\chi^2$ ($K_{\chi^2} =5,000$) &  \multirow{3}[2]{*}{15.8} &\multirow{3}[2]{*}{\textbf{1.10\%}} & \underline{0.563} & \underline{0.943} & \underline{3128} \\
     $\wh \calB_{0.9}(\cdot)$ based on $\chi^2$ ($K_{\chi^2} =20,000$)&  & & \textbf{0.669} & \textbf{0.955 }& 3340 \\
    $\wh \calB_{0.9}(\cdot)$ based on ECDF-R &      & & 0.484 & 0.898 & \textbf{2611} \\
         \hline
     GAM  & -&1.36\% & 0.342 & 0.907 & 3376 \\
     \hline
    SAR   & -&2.03\% & 0.238     & 0.823     & 4065 \\
\hline
SARX   & -&1.65\% & 0.238     & 0.812    & 3176 \\
    \hline
    \hline
    \end{tabular}%
  \label{tab:compare}%
\end{table}%

For each testing day in 2019, we compute the mean absolute percentage error:
\[
{\rm MAPE} = {1 \over 48} \sum_{h=1}^{48} \frac{|\wh y_{t,h} - y_{t,h}|}{y_{t,h}}.
\]
Table \ref{tab:compare} summarizes the results. In terms of forecast
accuracy, PPC deliver the best performance with MAPE $=1.10\%$. Compared
to the GAM, SAR and SARX, this corresponds to the reduction in MAPE
of 19.1\%, 45.8\% and 33.3\%  respectively.
It is noticeable that  the coverage rate (CR) of the predictive bands are
smaller than the nominal level 0.9. One possible reason is
the small sample sizes of some groups,
for which the variances of noise are likely to be
underestimated, as the possible variation in the future is unlikely to be
reflected by the small number of the available observations. This is particularly
pronounced with the method based on ECDF-R.
Nevertheless PPC performs significantly better than the other method,
as all the three $\wh \calB_{0.9}(\cdot)$ listed in Table
\ref{tab:compare} provide much high
coverage rates (CR) for the whole
curve than GAM, SAR and SARX.
\begin{figure}
	\centering
	\includegraphics[width=4.7in,height=3in]{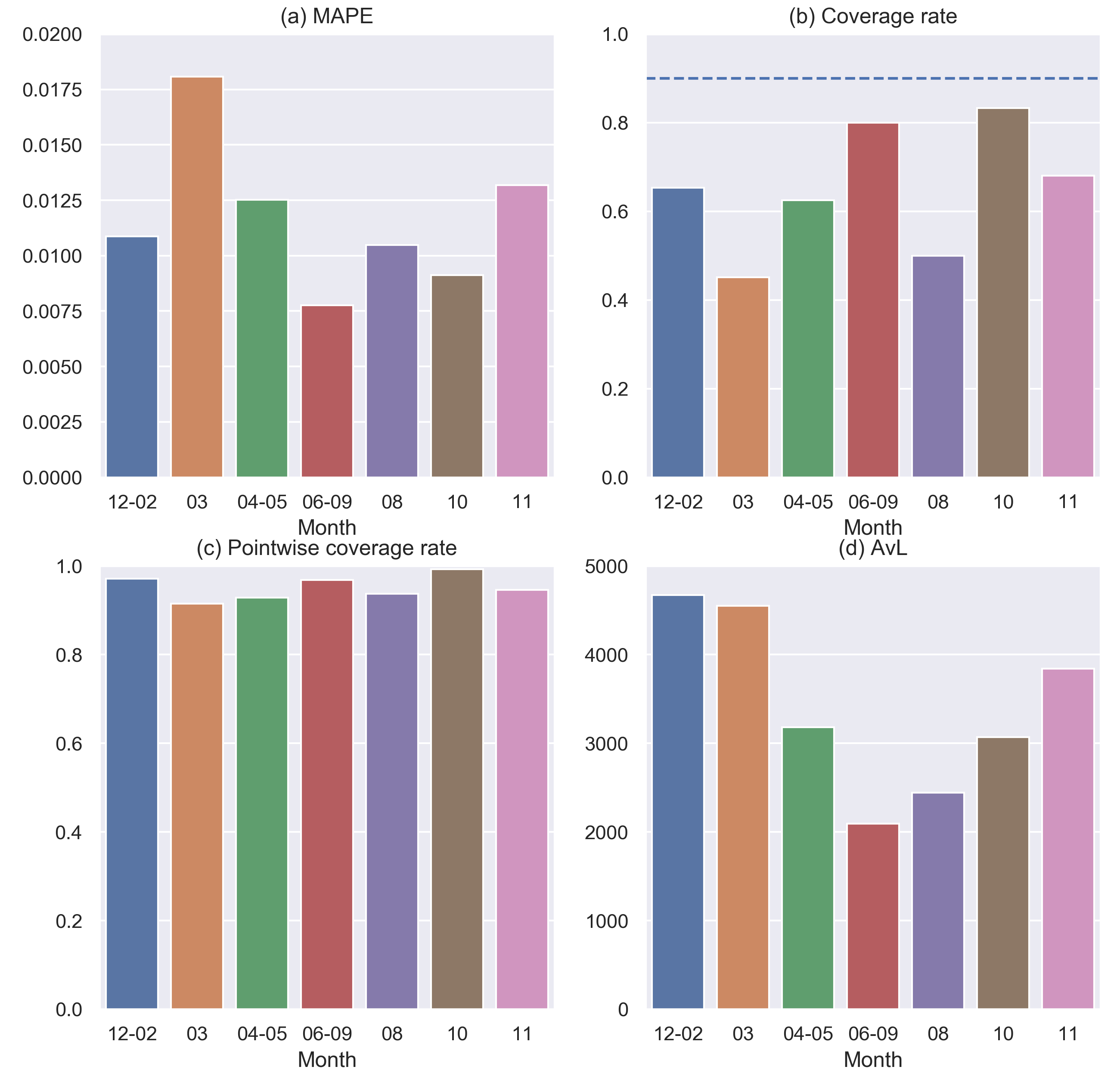}
\caption{\small Bar charts of MAPE, coverage rate, pointwise coverage rate
and AvL of  $\chi^2$-based $\wh \calB_{0.9}$ ($K_{\chi^2}$ = 20,000) over different months.}
	\label{fig:month}
\end{figure}
\begin{figure}
	\centering
	\includegraphics[width=4.7in,height=3in]{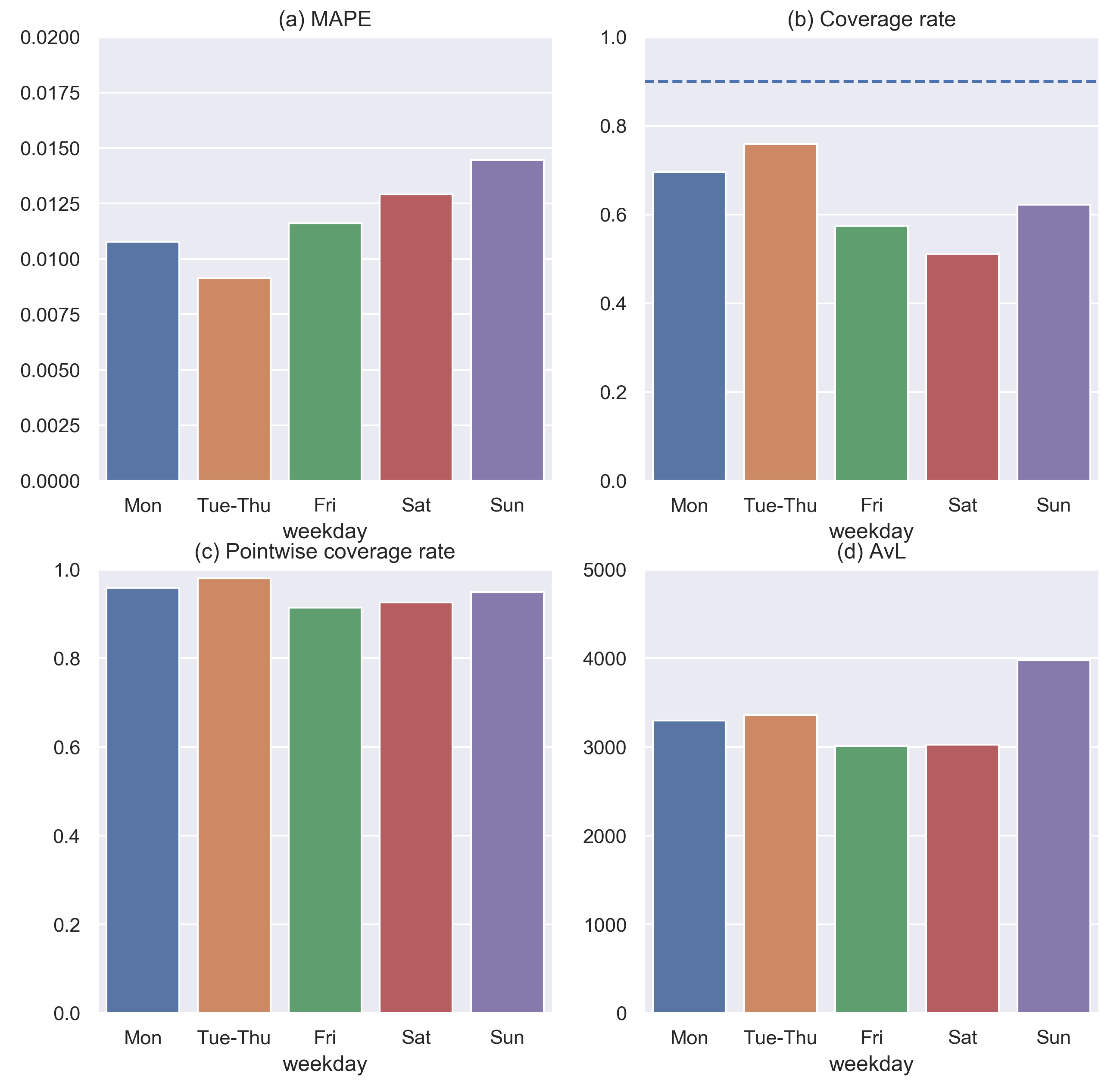}
\caption{\small Bar charts of MAPE, coverage rate, pointwise coverage rate
and AvL of  $\chi^2$-based $\wh \calB_{0.9}$  ($K_{\chi^2}$ = 20,000) over different weekdays.}
	\label{fig:weekday}
\end{figure}

One may argue that the comparison above is unfair as the predictive intervals based on
GAM, SAR and SARX are constructed in the pointwise manner rather than for a whole
curve.
Table \ref{tab:compare} also lists the pointwise coverage rates (PCR) of the different
methods for the 48 points (corresponding to the 48 30-minute intervals) on a daily curve.
It is interesting to observe that the two $\wh \calB_{0.9}(\cdot)$ based
on $\chi^2$ approximation provide significantly higher PCR than
those of GAM, SAR and SARX, and $\wh \calB_{0.9}(\cdot)$ based ECDF-R offers
comparable PCR to GAM with much smaller AvL.
It is worth to point out that
$\wh \calB_{0.9}(\cdot)$ based on $\chi^2$ with $K_{\chi^2} = 5, 000$ 
outperforms GAM, SAR and SARX in terms of all the three measures: delivering
higher CR (increases by 0.221 - 0.325), higher PCR (improves by 0.036 - 0.131) and narrower AvL (decreases by 48 - 937).

To appreciate the variation in forecasting performance over different
weekdays and different months,
Figures \ref{fig:month} and \ref{fig:weekday} display the bar-charts of
MAPE, CR, PCR and AvL of $\wh \calB_{0.9}(\cdot)$ based on
$\chi^2$-approximation with $K_{\chi^2} = 20,000$.
It is clear that the forecasting in summer is more accurate than that in winter
in terms of both MAPE (Figure \ref{fig:month}(a)) and AvL (Figure \ref{fig:month}(d)),
and the forecasting in Monday -- Friday is more accurate than that in weekends
in terms of MAPE (Figure \ref{fig:weekday}(a)). There is no clear pattern in terms
of the two coverage rates.


\begin{figure}[!t]
	\begin{subfigure}{.5\textwidth}
		\centering
		\includegraphics[width=1\linewidth]{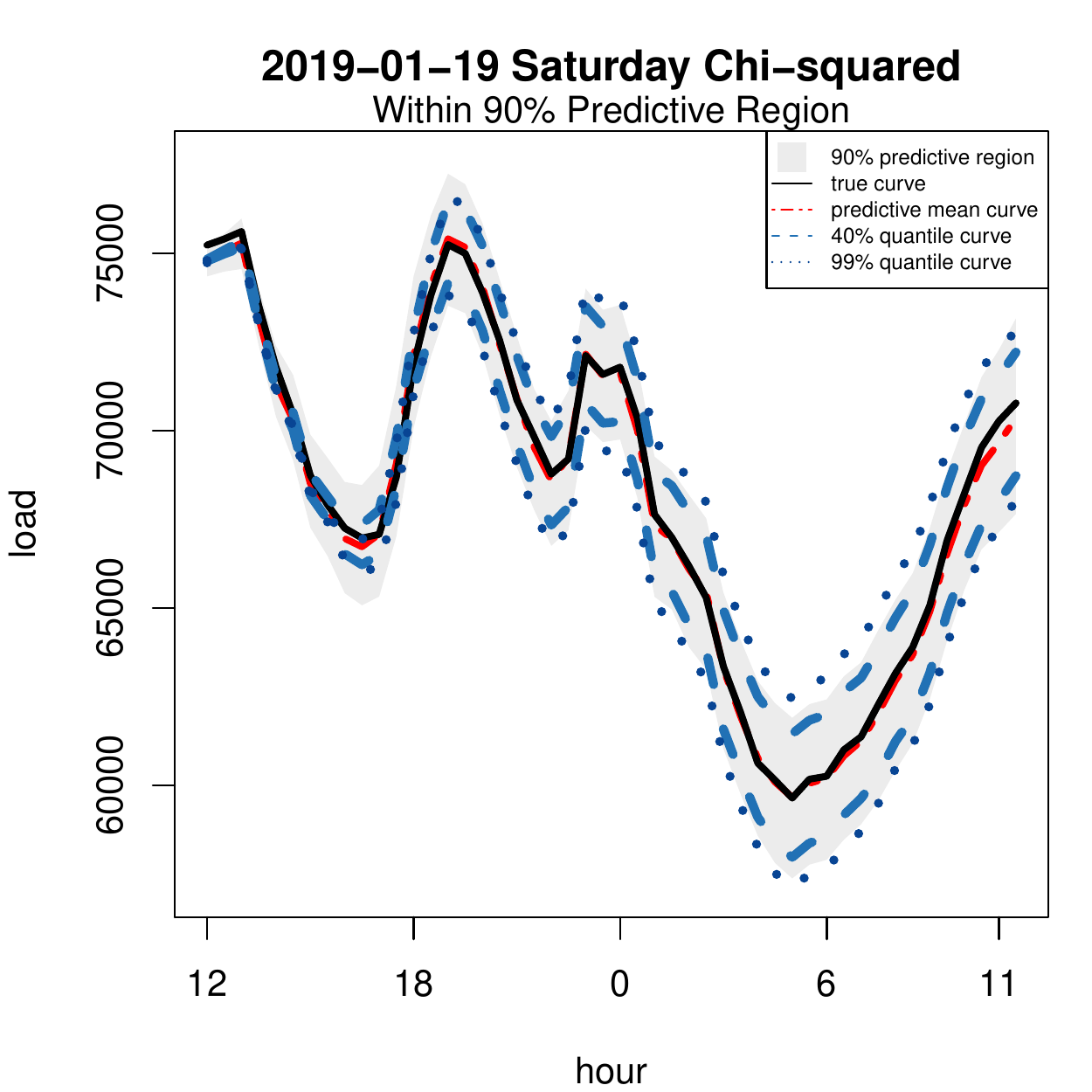}
	\end{subfigure}
	\begin{subfigure}{.5\textwidth}
		\centering
		\includegraphics[width=1\linewidth]{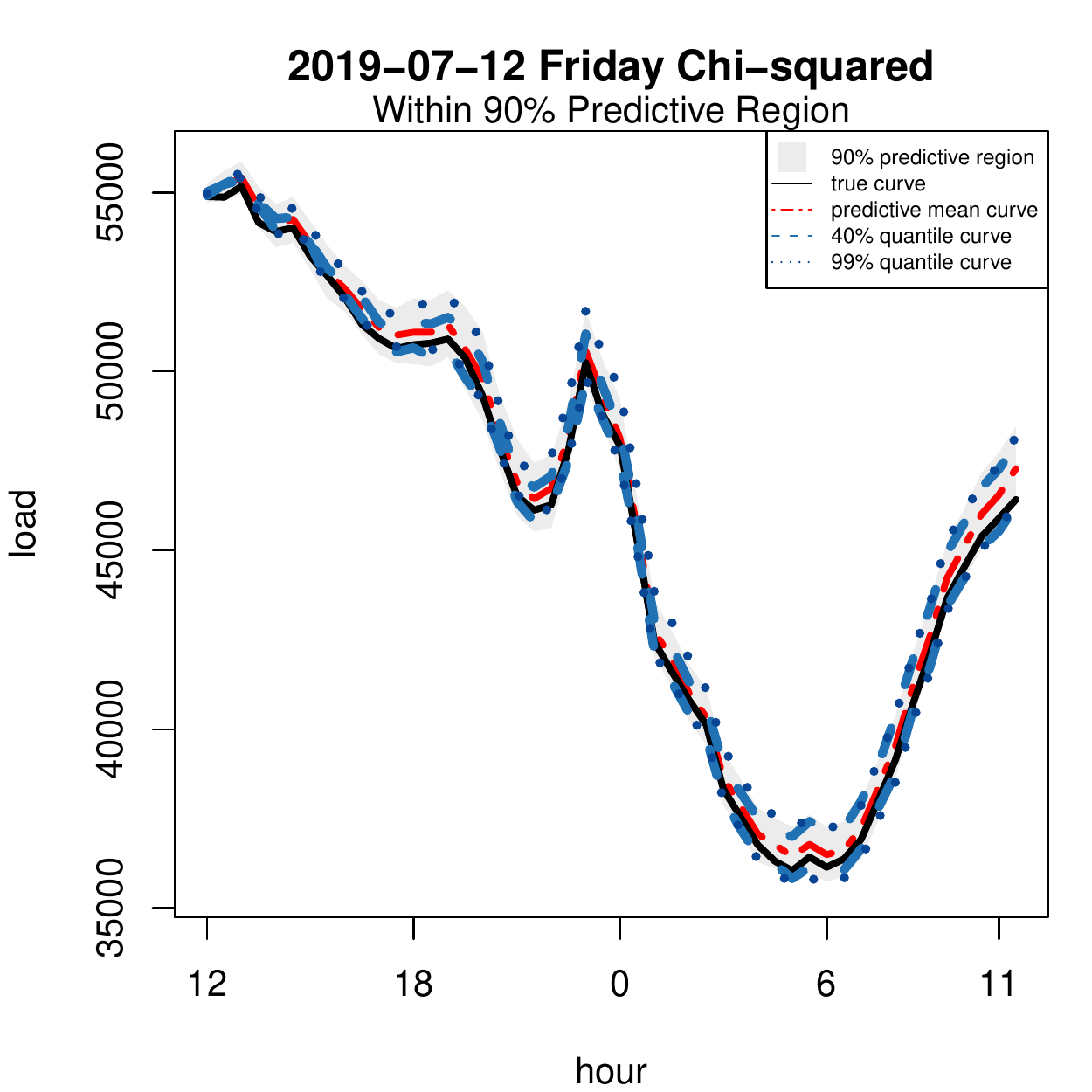}
	\end{subfigure}
	\caption{Plots of the true load curves (solid lines), predictive mean curves (dash-dotted lines), 90\% predictive regions (grey shadow areas) and predictive quantile curves at 99\% and 40\% confidence level (dotted and dashed lines) for 2 randomly selected curves, one from summer time and one from winter time.}
	\label{fig:quantiles}
\end{figure} 

Moreover, Figure \ref{fig:quantiles} elaborates 2 randomly selected forecasts, showing the predictive mean curves, $90\%$ predictive regions $\wh \calB_{0.9}(\cdot)$ and the quantile curves at 99\% and 40\% confidence levels in 2019 based on $\chi^2$-approximation with $K_{\chi^2} = 20,000$.   As expected, the predictive quantile curves at a higher confidence level is more ``outside'' than those at a lower level. The 99\% predictive quantile curves help to visualize the possible extreme cases, which is useful for the optimal controlling of the electricity operation system.  


\bigskip

\section{Conclusion} \label{sec:conclusion}
In this paper, we propose a novel method to construct three types of probabilistic predictors for curves (PPC): the predictive set, the predictive band and the predictive quantile with probability interpretation for daily electricity load
curves in a curve-to-curve linear regression framework. 
The PPC achieve excellent performance with coverage rates very close to the nominal probabilities for different simulated data generating processes. When applied to the French load curves, the proposed method provides much
smaller forecast errors, with almost half of that of the alternative
seasonal autoregressive models. Compared to the powerful generalized
additive model, it produces higher coverage rate with narrower average
length of the predictive regions. The significant improvement is likely due to
the curve regression setting which embeds the non-stationary daily patterns into
a stationary framework.
The constructed predictive intervals and the predictive quantile curves are meaningful and can be used in the future for risk hedging in the electricity management system.

%
%
%
%

%

\bibliographystyle{apalike}
\bibliography{Bibliography}

\end{document}